\documentclass[11pt]{article}
\usepackage{lmodern}
\usepackage{natbib}
\usepackage[a4paper, total={5.5in, 9in}]{geometry}
\usepackage{wrapfig}
\usepackage{amssymb,amsmath}
\usepackage{algorithm,algpseudocode}
\AtBeginDocument{}
\usepackage{sectsty}
\allsectionsfont{\centering}
\usepackage{titlesec}
\usepackage{booktabs}
\titlelabel{\thetitle.\quad}
\usepackage{microtype} 
\usepackage{float}
\usepackage{filecontents}
\providecommand{\keywords}[1]{\textbf{\textit{Keywords---}} #1}
\usepackage{enumitem}   
\usepackage{adjustbox}
\usepackage{xurl}
\usepackage{multirow}
\DeclareMathOperator{\tr}{Tr}
\usepackage{subcaption}
\usepackage[table]{xcolor}
\definecolor{Gray}{gray}{0.9}
\newcommand*{\angelaRev}[1]{\textcolor{black}{#1}}
\usepackage{titling}
\usepackage{authblk}

\title{Enhanced hyperalignment via spatial prior information}

\author[1]{Angela Andreella\footnote[1]{Corresponding author. \\ E-mail address: angela.andreella@unive.it}}
\affil[1]{Department of Economics, Ca' Foscari University of Venice,Venice, Italy}
\date{}
\author[2]{Livio Finos}
\affil[2]{Department of Developmental Psychology and Socialization, University of Padova, Padova, Italy}

\author[3]{Martin A Lindquist}
\affil[3]{Department of Biostatistics, Johns Hopkins Bloomberg School of Public Health}
\begin{document}
\maketitle
\begin{abstract}
Functional alignment between subjects is an important assumption of functional magnetic resonance imaging (fMRI) group-level analysis. However, it is often violated in practice, even after alignment to a standard anatomical template. \emph{Hyperalignment}, based on sequential \emph{Procrustes} orthogonal transformations, has been proposed as a method of aligning shared functional information into a common high-dimensional space and thereby improving inter-subject analysis. Though successful, current \emph{hyperalignment} algorithms have a number of shortcomings, including difficulties interpreting the transformations, a lack of uniqueness of the procedure, and difficulties performing whole-brain analysis. To resolve these issues, we propose the ProMises (Procrustes von Mises-Fisher) model. We reformulate functional alignment as a statistical model and impose a prior distribution on the orthogonal parameters (the von Mises-Fisher distribution). This allows for the embedding of anatomical information into the estimation procedure by  penalizing the contribution of spatially distant voxels when creating the shared functional high-dimensional space. Importantly, the transformations, aligned images, and related results are all unique. In addition, the proposed method allows for efficient whole-brain functional alignment. In simulations and application to data from \angelaRev{four} fMRI studies we find that ProMises improves inter-subject classification in terms of between-subject accuracy and  interpretability compared to standard \emph{hyperalignment} algorithms. 
\end{abstract}


\section{Introduction}
Multi-subject functional magnetic resonance imaging (fMRI) data analysis is important as it allows for the identification of shared cognitive characteristics across subjects. However, to be successful these analysis must properly account for individual brain differences. Indeed, it has been shown that the brains' anatomical and functional structures show great variability across subjects, even in response to identical sensory input \citep{Watson, Tootell, Hasson}. Various approaches have been proposed to deal with anatomical misalignment; these approaches align the images with a standard anatomical template \citep{Tal, Fischl, jenkinson2002improved}. However, these methods do not take into consideration the functional characteristics of the data; they fail to capture the shared functional response across subjects, ignoring the between-subject variability in the anatomical positions of functional loci. 

The problem of functional variability between subjects has long been known to neuroscientists \citep{Watson, Tootell, Hasson}. Indeed, this variation remains even after initial spatial normalization has been performed as a data preprocessing step. 
This can have serious consequences on group-level fMRI analysis where it is generally assumed that voxel locations are consistent across subjects after anatomical alignment \citep{lindquist2008statistical}. A lack of functional alignment can lead to erroneous statistical inference, resulting in both power-loss and reduced predictive accuracy \citep{wang2021bayesian}. 
To address this issue, \cite{Haxby} proposed a functional alignment technique called \emph{hyperalignment}, which uses orthogonal linear transformations to map brain images into a common abstract high-dimensional space that represents a linear combination of each subjects' voxel activation profile. In practice, \emph{hyperalignment} is a sequential application of the \emph{Procrustes} transformation \citep{Schoeman}, which consists of finding the optimal rotation/reflection that minimizes the distance between subjects' activation profiles. The abstract high-dimensional space created using \emph{hyperalignment} represents the common information space across individuals as a mixture of overlapping, individual-specific topographic basis functions \citep{haxby2020hyperalignment}. Individual-specific and shared functional information is modeled via high-dimensional transformations rather than transformations that rely on three-dimensional ($3D$) anatomical space. This innovative method of addressing the variability in the spatial location of functional loci across subjects has led to \angelaRev{promising new} research aimed at fostering an understanding of individual and shared cortical functional architectures \citep{HaxbyConn, haxby2020hyperalignment}.

Nevertheless, the approach has some shortcomings that remain to be addressed. First, \emph{hyperalignment} remixes data across spatial loci \citep{Haxby}. Therefore, its use in aligning data from the entire cortex may be questionable because it combines information from distant voxels to create the common abstract high-dimensional space \citep{haxby2020hyperalignment}. This potentially undermines the ability to  properly interpret the results. The method is powerful for classification; however, aligned images do not have a clear \angelaRev{topographical} interpretation. For this reason, \emph{hyperalignment} is applied more appropriately to a region of interest (ROI). An alternative is \emph{searchlight hyperalignment} \citep{Guntupalli}. Here overlapping transformations are calculated for overlapping searchlights in each subject and then aggregated into a single whole-cortex transformation. This ensures that the voxels of the aligned images are generated from a circumscribed ROI, and thus allows for a \angelaRev{topographical} interpretation of the final map. However, this final transformation is no longer an orthogonal matrix, and therefore does not preserve the content of the original data, namely the similarity/dissimilarity in the response between pairs of voxels. In addition, the searchlights are imposed a priori and do not allow voxels outside of the predefined search radius to influence the estimation process.
Second, as we show later in this work, the solutions calculated using \emph{hyperalignment} are not unique in the sense that they depend directly on the order in which individuals are entered into the algorithm, as it is a sequential version of \emph{generalized Procrustes analysis} (GPA) \citep{GowerGPA}. Note that even the original GPA does not provide a unique solution. Third, while the idea of applying functional alignment to fMRI data is important, its application to whole-brain analysis is problematic. In fact, most \emph{Procrustes}-based methods are based on singular value decomposition of square matrices whose dimensions equal the number of voxels. Therefore, it is infeasible to compute when working with the dimensions commonly used in whole-brain fMRI studies.

This paper proposes an approach that uses \emph{hyperalignment} as a foundational principle, but at the same time resolves the aforementioned outstanding issues. The proposed method allows a researcher to decide how to combine individual responses to construct the common abstract high-dimensional space where the shared functional information is represented. This is possible because the objective function of the Procrustes problem can be considered as a least-squares problem. We reformulate it as a statistical model, which we denote the \emph{ProMises} (Procrustes von Mises-Fisher) model. We assume a probability distribution for the error terms as well as a prior distribution for the orthogonal matrix parameter to restrict the range of possible transformations used to map the neural response into the common abstract high-dimensional space. The constraint is based on specifics that the researcher has defined inside the hyperparameter of the prior distribution. We explain how to define this hyperparameter in an appropriate manner. Through a simple formulation, we show how the proposed model allows \angelaRev{topographical} information to be inserted into the estimation process and simultaneously computes unique solutions. Using the proposed model, researchers can move away from a black-box approach, and instead better understand how functional alignment works by providing a neurophysiological interpretation of the aligned images as well as related results. In this way, the local radial constraints used in \emph{searchlight hyperalignment} are surpassed. The model can incorporate them directly into the Procrustes estimation process, thus retaining all of \emph{hyperalignment}'s intrinsic properties, such as preserving the vector geometry. 

The solution provides a unique representation of the aligned images and the related transformations (e.g., classifier coefficients, statistical tests, and correlations) in standardized anatomical brain space. In addition, the idiosyncratic topographies encoded inside the orthogonal transformation and the shared functional information do not depend on the specific reference matrix used by the algorithm. On the contrary, given that \emph{hyperalignment} is a sequential approach of the Procrustes problem, the reference is not clear; it depends on the order of the subjects and the algorithm's successive steps. In our model, the coefficients forming the basis function of the common abstract high-dimensional space are unique and reflect the orientation's prior information. Finally, to allow for whole-brain analysis, we propose a computationally efficient version of the ProMises model, where proper semi-orthogonal transformations project these square matrices into a lower-dimensional space without loss of information.

The paper is organized as follows. Subsection \ref{Procrustes} outlines functional alignment via \emph{Procrustes}-based methods, \angelaRev{while Subsection \ref{hyp-related} describes some methods in the literature related to them, emphasizing their weaknesses.}
Thereafter, we offer a solution to these problems in Subsection \ref{ProMises}, introducing the ProMises model as well as an efficient version of the model for whole-brain analysis. Subsection \ref{data} describes the \angelaRev{four} datasets explored. Finally, Section \ref{experiments} illustrates the performance of the proposed alignment method within a multi-subject classification framework. We compare the results with those obtained using no functional alignment (i.e, anatomical alignment only  \citep{jenkinson2002improved}) and functional alignment (after anatomical alignment) using GPA \citep{GowerGPA} and \emph{hyperalignment} \citep{Haxby}.

\section{Methods}
\subsection{Functional alignment by \emph{Procrustes}-based method}\label{Procrustes}

The group neural activation can be described by a set of matrices, $\{\boldsymbol{X}_i \in \mathbb{R}^{t \times v}\}_{i = 1, \dots, m}$, one for each subject $i$. Here the $t$ rows represent the response activation of $v$ voxels at each time point, and the $v$ columns represent the time series of activation for each voxel. The rows are ordered consistently across all subjects because the stimuli are time-synchronized; however, the columns are not assumed to correspond across subjects \citep{Watson, Tootell, Hasson}. The functional alignment step is thus crucial for consistently comparing activation in a certain voxel between subjects \citep{haxby2020hyperalignment}.

The most famous method for assessing the distance between matrices is the \emph{Procrustes} transformation \citep{Gower}. In simple terms, it uses similarity transformations (i.e., rotation and reflection) to match matrice(s) onto a target matrix as close as possible according to the Frobenius distance, using least-squares techniques.

When one matrix, $\boldsymbol{X}_i$, is transformed into the space of another $\boldsymbol{X}_j$ via orthogonal transformation $\boldsymbol{R} \in \mathcal{O}(v)$, where $\mathcal{O}(v)$ defines the set of orthogonal matrices in $\mathbb{R}^{v \times v}$, the \emph{Procrustes} problem is called the \emph{orthogonal Procrustes problem} (OPP):
\begin{equation}
\label{eq:OPP}
\min_{\boldsymbol{R} \in \mathcal{O}(v)} || \boldsymbol{X}_i \boldsymbol{R} - \boldsymbol{X}_j ||_{F}^2,
\end{equation}
where $|| \cdot ||_{F}$ denotes the Frobenius norm. The minimum is given by $\boldsymbol{R}= \boldsymbol{U} \boldsymbol{V}^\top$, where $\boldsymbol{U}$ and $\boldsymbol{V}$ come from the singular value decomposition (SVD) of $\boldsymbol{X}_i^\top \boldsymbol{X}_j = \boldsymbol{U} \boldsymbol{\Sigma} \boldsymbol{V}^\top$ \citep{Peter}.

Generally, fMRI group-level analysis deal with $m \ge 2$ subjects. In this case, the functional alignment can be based on the GPA \citep{GowerGPA}:
\begin{align}
\label{eq:GPA}
\min_{\boldsymbol{R}_i \in \mathcal{O}(v)} \sum_{i=1}^{m} || \boldsymbol{X}_i \boldsymbol{R}_i - \boldsymbol{M} ||_{F}^2,
\end{align}
where $\boldsymbol{M}$ is the element-wise arithmetic mean of transformed matrices $\boldsymbol{X}_i \boldsymbol{R}_i$, also called the \emph{reference} matrix. Equation \eqref{eq:GPA} does not have a closed form solution, and is solved using an iterative procedure proposed by \cite{Gower}. Alternatively, \emph{hyperalignment} \citep{Haxby} can be used, which is based on the sequential use of the OPP defined in Equation \eqref{eq:OPP}.  

Importantly, both GPA and \emph{hyperalignment} appear to have some shortcomings to resolve in order to yield unique, reproducible, and interpretable results.
First, the orthogonal transformation $\boldsymbol{R}_i$, computed via these methods, can combine information from every voxel inside of the cortical field or ROI. Anatomical structure is ignored, implicitly assuming that functional areas can incorporate neural activation of voxels from any part of the cortical area. A solution commonly used in the field is the searchlight approach proposed by \cite{Guntupalli}. However, this method assumes an optimal searchlight size, which must be defined by the researcher, thus introducing some degree of arbitrariness.     \angelaRev{Another approach which is more efficient than \emph{searchlight hyperalignment} is to cluster voxels into sets of sub-regions across the whole brain as discussed by \cite{bazeille2021empirical}. In this way, the non-orthogonality problem of the \emph{searchlight hyperalignment} approach is surpassed, but the set of sub-regions must again be defined a priori. In addition, in the parcel boundaries the optimality of this approach is not assured.} Second, both methods return more than one solution (i.e., GPA has an infinite set of solutions, while \emph{hyperalignment} has $m!$ solutions, where $m$ is the number of subjects). The $\boldsymbol{R}_i$ computed via GPA are unique up to rotations. Instead, the $\boldsymbol{R}_i$ calculated via \emph{hyperalignment} strictly depends on the order of the subjects entering the algorithm. Being a sequential approach of the OPP, the choice of  \emph{reference} matrix is not clear, and every matrix used as a starting matrix leads to different common high-dimensional spaces.

\angelaRev{\subsection{\emph{Hyperalignment}-related method}\label{hyp-related}
\angelaRev{After \cite{Haxby}, various modifications of \emph{hyperalignment} have appeared in the literature. We do not list all the possible modifications here, and for a complete review please see \cite{cai2020incorporating, bazeille2021empirical}. One of the most successful methods in the literature is the Shared Response Model (SRM) proposed by \cite{Xu}, which is a probabilistic model that computes a reduced dimension shared feature space. The method was also reformulated in matrix format by \cite{shvartsman2018matrix} and later analyzed by \cite{cai2020incorporating, bazeille2021empirical}. In short, SRM estimates a semi-orthogonal matrix with dimensions $v \times k$, where $k$ is a tunable hyper-parameter representing the number of shared features. Therefore, as discussed by the authors, SRM returns a non unique set of semi-orthogonal transformations that leads to the loss of: (1) the original spatial characteristics; and (2) the topographical interpretation of the final aligned data. Similar to \emph{hyperalignment} and GPA, SRM does not allow for the incorporation of spatial anatomical information into the estimation process, unlike the proposed ProMises model. \cite{Xu}'s method improves upon \emph{hyperalignment} in terms of classification accuracy and scalability, but it analyzes the first $k$ dimensions (i.e., latent variables), while \emph{hyperalignment} is not constructed to be a reduction dimension technique.} }

\angelaRev{There are several promising functional alignment approaches in the literature that are not based on Procrustes theory, such as the optimal transport approach proposed by \cite{bazeille2019local}. Comparisons with these methods would be interesting, but in this paper we limit ourselves to analyzing Procrustes-based approaches such as GPA and \emph{hyperalignment}.}

\subsection{ProMises model}\label{ProMises}

The focus of this paper is to resolve the non-uniqueness and mixing problem of the transformations computed via GPA and \emph{hyperalignment}. Indeed, solving these issues allows for the exploration of the structural neuroanatomy of functionally aligned matrices and their related transformations. The ProMises model resolves both points in an elegant and simple way, defining a hyper parameter for tuning the locality constraint. We stress here that it computes a unique solution that preserves the fine-scale structure and allows for penalization of spatially distant voxels in the construction of the shared high-dimensional space, assuming that the anatomical alignment is not too far from the central tendency. 

To achieve these goals we seek to insert prior information about the structure of $\boldsymbol{R}_i$ into Equations \eqref{eq:OPP} - \eqref{eq:GPA}, which converts the set of possible orthogonal transformation solutions to a unique solution reflecting the prior information embedded. This is possible if we analyze the \emph{Procrustes} problem from a statistical perspective. In short, the least squares problem formulate in Equations \eqref{eq:OPP} - \eqref{eq:GPA} are reformulated as a statistical model, which allows for the definition of a prior distribution on $\boldsymbol{R}_i$. To be precise, the \angelaRev{the difference between $\boldsymbol{X}_i \boldsymbol{R}_i$ and $\boldsymbol{M}$} described in \eqref{eq:GPA} can be viewed as an error term that is assumed to be normally distributed in our statistical model defined in the following subsection.

\subsubsection{Model}

The minimization problem defined in Equation \eqref{eq:GPA} can be reformulated as follows:
\begin{align}
\label{eq:eq6}
\boldsymbol{X}_i=  \boldsymbol{M} \boldsymbol{R}_i^\top + \boldsymbol{E}_i \quad\quad \text{subject to }\quad \boldsymbol{R}_i \in \mathcal{O}(v),
\end{align}
where $\boldsymbol{E}_i \in \mathbb{R}^{t \times v}$ is the error matrix to minimize and $\boldsymbol{M} \in \mathbb{R}^{t \times v}$ is the \emph{reference} matrix. 

We assume a multivariate normal matrix distribution \citep{Gupta} for the error terms $\boldsymbol{E}_i$. Each row of $\boldsymbol{E}_i$ is distributed as a multivariate normal distribution with mean $0$ and covariance $\boldsymbol{\Sigma}_v$. The observed data matrix $\boldsymbol{X}_i$ is then described as a random Gaussian perturbation of $\boldsymbol{M}$. The rotation matrix parameter $\boldsymbol{R}_i$ allows for the representation of each data matrix $\boldsymbol{X}_i$ in the shared functional space. In other words, the model simply reflects the assumption underlining \emph{hyperalignment}, namely that neural activity in different brains are noisy rotations of a common space \citep{Haxby}. In this paper, we assume $\boldsymbol{\Sigma}_v = \boldsymbol{I}_v$, where $\boldsymbol{I}_v$ is the identity matrix of size $v$. The extension to an arbitrary type of variance matrix $\boldsymbol{\Sigma}_v$ and incorporation of its estimation into the ProMises model is discussed in \cite{Andreella}.

\subsubsection{Prior information}

Rephrasing the \emph{Procrustes} problem as a statistical model allows us to impose a prior distribution on the orthogonal parameter $\boldsymbol{R}_i$. With the constraint $\boldsymbol{R}_i \in \mathcal{O}(v)$ in \eqref{eq:eq6}, the probability distribution for $\boldsymbol{R}_i$ must take values in the Stiefel manifold $V_{v}(\mathbb{R}^v)$ (i.e., the set of all $v$-dimensional orthogonal bases in $\mathbb{R}^{v}$). An attractive distribution on $V_{v}(\mathbb{R}^v)$ is the matrix von Mises-Fisher distribution, introduced by \cite{Downs} and further investigated by many others \citep{Khatri,Prentice,Chikuse1,Chikuse,MardiaB}. It is defined as follows:
\begin{align}
f(\boldsymbol{R}_i) = C(\boldsymbol{F},k) \exp\Big\{\tr(k \boldsymbol{F}^T \boldsymbol{R}_i) \Big\},
\end{align}
where $\tr(\cdot)$ defines the trace of a square matrix (i.e., the sum of elements on the main diagonal), \(C(\boldsymbol{F},k)\) is a normalizing constant, $k \in \mathbb{R}_{\ge0}$ is the concentration parameter, and $\boldsymbol{F} \in \mathbb{R}^{v \times v}$ is the location matrix parameter. 

The parameter $k$ balances the amount of concentration of the distribution around $\boldsymbol{F}$. As $k \rightarrow 0$, the prior distribution approaches a uniform distribution, representing the unconstrained case. In contrast, as $k \rightarrow +\infty$, the prior tends toward a distribution concentrated at a single point, representing the maximum constraint.

The polar part of $\boldsymbol{F}$ represents the mode of the distribution, and is unique if and only if $\boldsymbol{F}$ has full rank \citep{Jupp}. In addition, the matrix von Mises-Fisher distribution is a conjugate prior (i.e., the posterior distribution has closed-form expression in the same family of distributions as the prior) for the matrix normal distribution with posterior parameter equal to $\boldsymbol{X}_i^\top \boldsymbol{M} + k \boldsymbol{F}$. The solution for $\boldsymbol{R}_i$ is unique if and only if $\boldsymbol{X}_i^\top \boldsymbol{M} + k \boldsymbol{F}$ has a full rank.  
Therefore, in the following, we define $\boldsymbol{F}$ such that it is of full rank and incorporates valuable information about the final high-density common space.

The elements of the final high-dimensional common space are composed of linear combinations of voxels \citep{haxby2020hyperalignment}. 
Thus, $\boldsymbol{F}$ can be properly defined such that these combinations emphasize nearby voxels and penalize distant voxels. In this way, the anatomical structure of the cortex is used as prior information in the estimation of $\boldsymbol{R}_i$. The idea is that nearby voxels should have similar rotation loadings, whereas voxels that are far apart should have less  similar loadings. The hyperparameter $\boldsymbol{F}$ is defined as a Euclidean similarity matrix using the $3D$ anatomical coordinates of $x$, $y$, and $z$ of each voxel:
\begin{align}\label{euclidean}
\boldsymbol{F} =  [\exp\{-d_{ij}\}] = \Big[\exp\Big\{- \sqrt{(x_i - x_j)^2 + (y_i - y_j)^2 + (z_i - z_j)^2}\Big\}\Big],
\end{align}
where $i,j = 1,\dots v$. In this way, $\boldsymbol{F}$ is a symmetric matrix with ones in the diagonal, which means that voxels with the same spatial location are combined with weights equal to $1$, and the weights decrease as the voxels to be combined become more spatially distant. 

$\boldsymbol{F}$ can also be specified via geodesic distances to exploit the intrinsic brain curve structure, or via the \angelaRev{Dijkstra} distance \citep{dijkstra1959note} if surface-based data are analyzed. In addition, another type of Minkowski measure \citep{upton2014dictionary} may be used; however, it is important to carefully consider the type of spatial information available, i.e., the distance used must be reasonable in the context of the data. In this case, the Euclidean similarity matrix is an attractive measure for detecting how close the voxels overlap. In addition, $\boldsymbol{F}$, defined as in Equation \eqref{euclidean}, has full rank, which is a necessary condition for having a unique solution for $\boldsymbol{R}_i$. The graphical representations of the Euclidean distance matrix using the $3D$ anatomical coordinates of voxels (or vertices of a surface grid) are proposed in the next section analyzing two types of datasets (face and object recognition and movie watching).


In summary, the ProMises model returns a solution that is a slight modification of the OPP solution that \cite{Peter} developed for the case of two subjects, as well as a slight modification of the GPA solution that \cite{GowerGPA} proposed for the case of multiple subjects. The modification is based on applying the SVD to $\boldsymbol{X}_i^\top \boldsymbol{M} +k \boldsymbol{F}$ instead of $\boldsymbol{X}_i^\top \boldsymbol{M}$. Thus, prior information about $\boldsymbol{R}_i$ enters the SVD step through the specification of $\boldsymbol{F}$, with the term $k$ balancing the relative contribution of $\boldsymbol{X}_i^\top \boldsymbol{M}$ and $\boldsymbol{F}$. Thanks to this regularization, the ProMises model returns a set of unique transformations that correspond to the anatomical brain structure, exploiting the spatial location of voxels in the brain, or ROI. Hence, the ability to define the parameter $\boldsymbol{F}$ guarantees a \angelaRev{topographical} interpretation of the results, as we will see in the next section.

\subsubsection{Efficient ProMises model}
The ProMises model returns a unique orthogonal transformation for each subject; however, it cannot be applied to the entire brain due to the extensive computational burden. This is due to the fact that at each step we must compute $m$ singular value decompositions of $v \times v$ matrices leading to polynomial time complexity.

To allow for whole-brain analysis, we propose the Efficient ProMises model, which allows for a faster functional alignment without loss of information. In practice, the Efficient ProMises model projects matrices $\boldsymbol{X}_i$ into a $t$ lower-dimensional space via specific semi-orthogonal transformations $\boldsymbol{Q}_i \in \mathbb{R}^{v \times t}$ \citep{Abadir, Grob} which preserve all of the information in the data. It aligns the reduced $t\times t$ matrices $\{\boldsymbol{X}_i \boldsymbol{Q}_i \in {\rm I\!R}^{t \times t }\}_{i = 1, \dots, m}$, and back-projects the aligned matrices to the original $t\times v$-size matrices $\{\boldsymbol{X}_i \in {\rm I\!R}^{t \times v }\}_{i = 1, \dots, m}$ using the transpose of these semi-orthogonal transformations ($\boldsymbol{Q}_i^\top \in \mathbb{R}^{t \times v}$).

No loss of information occurs because the minimum of Equation \eqref{eq:GPA} using $\{\boldsymbol{X}_i \boldsymbol{Q}_i \in {\rm I\!R}^{t \times t }\}_{i = 1, \dots, m}$ is equivalent to the one obtained using the original data. This is due to the fact that the Procrustes problem analyzes the first $t \times t$ dimensions of $\boldsymbol{R}_i$. Hence, the minimum remains the same if we use as our semi-orthogonal matrices $\{\boldsymbol{Q}_i\}_{i = 1, \dots, m}$ the ones obtained from the thin singular value decomposition of $\{\boldsymbol{X}_i \in {\rm I\!R}^{t \times v }\}_{i = 1, \dots, m}$.

The algorithms describing the ProMises model estimation process and its Efficient version are reported in \ref{pseudocode}. For further details and proofs about the ProMises model and its efficient version, please see \cite{Andreella}.

\subsection{fMRI Datasets}\label{data}

The performance of the proposed method is assessed using two fMRI datasets from \cite{Haxby} and one from \cite{haxby2001distributed} summarized in Table \ref{tab:tab1}. \angelaRev{We analyzed an additional dataset collected by \cite{duncan2009consistency}.} The datasets differ in several key characteristics, including number of subjects, whether data is extracted from an ROI or the whole-brain, and the number of time points, voxels and  stimuli. In addition, they differ depending on whether the data is in volumetric space or on a surface. These differences will allow us to evaluate the performance of the proposed model in a number of different circumstances. 

\begin{table}
\begin{adjustbox}{width=\columnwidth,center}
    \centering
    \begin{tabular}{@{}lrlrrr@{}}
        \textbf{Dataset} & \textbf{Subjects} & \textbf{ROI} & \textbf{Length} & \textbf{Voxels} & \textbf{Stimuli} \tabularnewline 
        \toprule
        \rowcolor{Gray}
        Faces and objects & 10 & Ventral temporal cortex & 56 & 3509 & 8 \tabularnewline
        Visual object recognition & 6 & Whole brain & 121 & \angelaRev{39912} & 8 \tabularnewline
                \rowcolor{Gray}

        Raiders & 31 & Ventral temporal cortex & 2662 & 883 & 400\tabularnewline
        Raiders & 31 & Occipital lobe & 2662 & 653 & 400\tabularnewline
                \rowcolor{Gray}

        Raiders & 31 & Early visual cortex & 2662 & 484 & 400\tabularnewline
      \angelaRev{  Words and objects} & \angelaRev{12} & \angelaRev{Whole brain} & \angelaRev{164} & \angelaRev{73574} & \angelaRev{5} \tabularnewline
        \bottomrule
    \end{tabular}
    \end{adjustbox}
    \caption{Description of the datasets used in our analysis. They differ in factors such as number of subjects, the region of interest, as well as the number of time points, voxels, and stimuli.}
    \label{tab:tab1}
\end{table}

\begin{figure}
\centering \includegraphics[width=.6\linewidth]{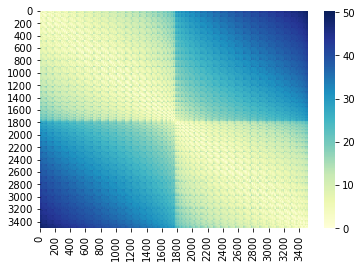}  
 \caption{\label{fig:figs2} Representation of the Euclidean distance matrix to compute location parameter $\boldsymbol{F}$ of the von Mises-Fisher distribution using the $3D$ coordinates of the voxels from the \emph{faces and objects} dataset.}
\end{figure}

The first dataset, referred to as \emph{faces and objects}, is a block-design fMRI study aimed at analyzing face and object representations in the human ventral temporal (VT) cortex. It is composed of fMRI images of $10$ subjects with eight runs per subject. In each run, subjects look at static, gray-scale images of faces and objects (i.e., human females, human males, monkeys, dogs, houses, chairs, and shoes). The subject views these images for $500$ ms with $1500$ ms inter-stimulus intervals. Each block consists of viewing $16$ images from one category, corresponding to a one-back repetition detection task for each subject. Blank intervals of $12$ divide the blocks. Each run contains one block of each stimulus category. Brain images were acquired using a $3$T Siemens Allegra scanner with a standard bird-cage head coil. Whole brain volumes of $32$ $3$ mm thick axial slices [TR = $2$ s, TE = $30$ ms, flip angle = $90^\circ$, $64 \times 64$ matrix, FOV = $192$ mm $\times 192$ mm] were obtained that included all of the occipital and temporal lobes and all but the most dorsal parts of the frontal and parietal lobes. High resolution T$1$-weighted images of the entire brain were obtained in each imaging session [MPRAGE, TR = $2.5$ s, TE = $4.3$ ms, flip angle = $8^\circ$, $256 \times 256$ matrix, FOV = $256$ mm $ \times 256$ mm, $172$ $1$ mm thick sagittal images]. For further details about the experimental design and data acquisition, please see \cite{Haxby}. Here, the analysis is focused on the $3509$ voxels within the VT cortex. Figure \ref{fig:figs2} shows the Euclidean distance matrix used to calculate the location matrix parameter $\boldsymbol{F}$ of the von Mises-Fisher distribution for this dataset. \angelaRev{In this analysis, we define $\boldsymbol{F}$ as a Euclidean similarity matrix; see Equation \eqref{euclidean}.} The two visible blocks represent the left and right VT cortex. A jump of four units \angelaRev{(i.e., voxel index $ijk$ units) in the $i^{th}$ dimension} exists between voxel $1782$ and voxel $1783$, corresponding to the corpus callosum. 

The second dataset, referred to as \emph{visual object recognition} has a similar structure as the \emph{faces and objects} dataset, where six subjects are viewing images of faces, cats, five
categories of man-made objects, and nonsense pictures for 500 ms with an inter-stimulus interval of 1500
ms. Brain images were acquired on a GE $3$T scanner (General Electric, Milwaukee, WI). Whole brain volumes of $40$ $3.5$ mm thick sagittal images [TR = $2500$ ms, , TE = $30$ ms, flip angle = $90^\circ$, FOV = $24$ cm] were obtained. High-resolution T$1$-weighted spoiled gradient recall (SPGR) images were obtained for each subject to provide detailed anatomy [$124$ $1.2$ mm thick sagittal images, FOV =$24$ cm]. For further details, see \cite{haxby2001distributed}. Here the analysis is focused on the use of whole-brain data consisting of \angelaRev{39912} voxels. \angelaRev{Having a large number of voxels, we here use the Efficient ProMises model to align the brain images. In this case, the location matrix parameter is a lower-dimensional version of $\boldsymbol{F} \in \mathbb{R}^{v \times v}$, i.e., the similarity Euclidean matrix defined in Equation \eqref{euclidean}. This new location matrix parameter must take values in $\mathbb{R}^{t \times t}$. It is expressed as $\boldsymbol{Q}_i^\top \boldsymbol{F} \boldsymbol{Q}_M$, where $\boldsymbol{Q}_i$ is the semi-orthogonal matrix coming from the thin singular value decomposition of $\boldsymbol{X}_i$, and $\boldsymbol{Q}_M$ is the semi-orthogonal matrix coming from the thin singular value decomposition of $\boldsymbol{M}$.}

\begin{figure}
\centering
\subfloat[Ventral temporal cortex]{\includegraphics[width = .33\textwidth]{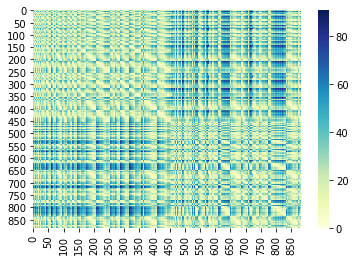}} 
\subfloat[Early visual cortex]{\includegraphics[width = .33\textwidth]{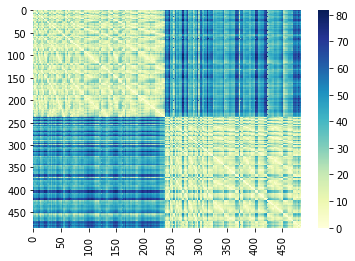}}
\subfloat[Occipital lobe]{\includegraphics[width = .33\textwidth]{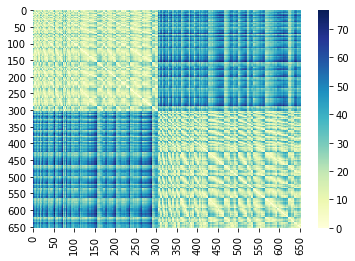}}
\caption{\label{fig:figs4} Representation of the Euclidean distance matrix used to compute the location parameter $\boldsymbol{F}$ of the von Mises-Fisher distribution using the $3D$ coordinates of the voxels from the \emph{raiders} dataset.}
\end{figure}

The third dataset, referred to as \emph{raiders}, consists of $31$ subjects watching the movie ``Raiders of the Lost Ark'' (1981). The movie session was split into eight parts of approximately $14$ minutes. Brain images were acquired using a 3T Philips Intera Achieva scanner with an eight-channel head coil. Brain volumes were obtained consisting of 41 3 mm thick sagittal images [R = 2.5 s, TE = 35 ms, Flip angle = 90°, 80 x 80 matrix, FOV = 240 mm x 240 mm]. High resolution T$1$ weighted images of the entire brain were obtained in each imaging session [MPRAGE, TR = $9.85$ s, TE = $4.53$ ms, flip angle = $8^\circ$, $256 \times 256$ matrix, FOV = $240$ mm $ \times 240$ mm, $160$ $1$ mm thick sagittal images]. For more details about subjects, MRI scanning parameters, data preprocessing, \angelaRev{and} ROI definition, see \cite{Haxby}. Here the analysis is focused on ROIs in the VT cortex ($883$ voxels), occipital lobe (LO; $653$ voxels), and early visual (EV; $484$ voxels) cortex. Figure \ref{fig:figs4} shows the Euclidean distance matrix using the $3D$ coordinates from the ROIS defined over the VT cortex, LO, and EV cortex. The two blocks represent the left and right parts of the ROIs. In this case, the $3D$ coordinates describe the vertices of a surface grid based on the cortex envelope. The mapping from the volume to the surface was computed using the FreeSurfer software \citep{Fischl}. \angelaRev{As in the first analysis (i.e., faces and objects), we defined the location matrix as a Euclidean similarity matrix as seen in Equation \eqref{euclidean}. However, in this case, the $3$-D coordinates refer to the vertices of the surface grid, as mentioned before. Thus, we could have defined $F$ using geodesic distances. However, we found no substantial improvement in the results. Therefore, we prefer to use Euclidean distance since it provides a full-rank matrix (i.e., a necessary property to achieve the uniqueness of the solution).}

\angelaRev{The fourth dataset, referred to as \emph{words and objects}, is a block-design fMRI study to analyze brain regions, such as occipital temporal cortex, associated with functional word and object processing. In this study, $49$ subjects view images of written words, objects, scrambled objects, and consonant letter strings for $350$ ms with a $650$ ms fixation cross at the beginning of each trial. The functional data were acquired with a gradient-echo EPI sequence [TR $= 3000$ ms; TE$= 50$ ms; FOV$= 192 \times 192$; matrix$= 64 \times 64$] giving a notional resolution of $3 \times 3 \times 3$ mm. A high-resolution anatomical scan was acquired [T1-weighted FLASH, TR$= 12$ ms; TE$= 5.6$ ms; $1$ mm3 resolution] for each subject. For further details, see \cite{duncan2009consistency}. As in the \emph{visual object recognition} data analysis, we apply the Efficient ProMises model to align the whole brain image composed of $73574$ voxels.}

\angelaRev{For all analyses, we consider the set $\mathcal{K} = \{1, 2, \dots, 100\}$ as a collection of possible values for the concentration parameter $k$. The optimal value is estimated by cross-validation as explained in the next section.}

\angelaRev{The aim is to classify stimulus-driven response patterns in a left-out subject based on response patterns in other subjects.} These patterns are described via a sequence of voxels that might express an activation at a specific time point. It is a vector in a high-dimensional space, where each dimension represents a local feature (i.e., a voxel). Using multivariate pattern classification (MVPC) \citep{haxby2012multivariate}, the patterns of neural activities are then classified by analyzing their variability during different stimuli \citep{OToole, Kriegeskorte2}. We clarify here that the registration to standard MNI space is part of the preprocessing step. So, all functional alignment approaches are applied after spatial alignment to MNI space. \angelaRev{We then evaluate the ProMises model in terms of across-subject decoding accuracy \citep{bazeille2021empirical} and  interpretation of the final aligned images. We stress here that \cite{bazeille2021empirical} found that SRM \citep{Xu} and optimal transport \citep{bazeille2019local} outperformed \emph{hyperalignment} and \emph{searchlight hyperalignment} \citep{Guntupalli} at the ROI level. However, our aim is to provide a clear statistical model for functional alignment that permits one to incorporate spatial anatomical information into the estimation process, thereby leading to an optimal unique orthogonal transformation rather than focus on improving the classification predictive accuracy. The ProMises model proposed is compared in terms of between-subjects predictive accuracy with GPA, and \emph{hyperalignment} methods as well as anatomical alignment. We did not consider other related approaches (e.g., SRM \citep{Xu}, optimal transport \citep{bazeille2019local}) since we decided to focus on Procrustes-based approaches (i.e., those that minimize an objective function with an  orthogonality constraint) and to show the related variability of these approaches. For a complete review of functional alignment methods, please refer to \cite{bazeille2021empirical,cai2020incorporating}.}

We have developed a Python \citep{Python} module, -- \texttt{ProMisesModel}---available at \url{https://github.com/angeella/ProMisesModel} in line with the Python \texttt{PyMVPA} \citep{PyMVPA} package. We also have created the \texttt{alignProMises} \texttt{R} \citep{R} package available at \url{https://github.com/angeella/alignProMises} based on the C++ language.

\section{Results}\label{experiments}

\subsection{Faces and objects}\label{faces-and-objects}

The protocol for evaluating the performance of the ProMises model directly follows the one used in \cite{Haxby}. We classify the patterns of neural activation using a support vector machine (SVM) \citep{Vapnik}. The between-subject classification is computed using leave-one-out subject cross-validation. To avoid the circularity problem \citep{Kriegeskorte}, the alignment parameters and the regularization parameter $k$ (i.e., the concentration parameter) are fitted in the leave-one-out run using a nested cross-validation approach. The performance metric used is the mean accuracy over leave-one-out subjects and leave-one-out runs. Note for each of the methods compared the input data is spatially normalized to MNI space \citep{jenkinson2002improved}.

\begin{figure}[!ht]
	\minipage{1\textwidth}
	\includegraphics[width=\linewidth]{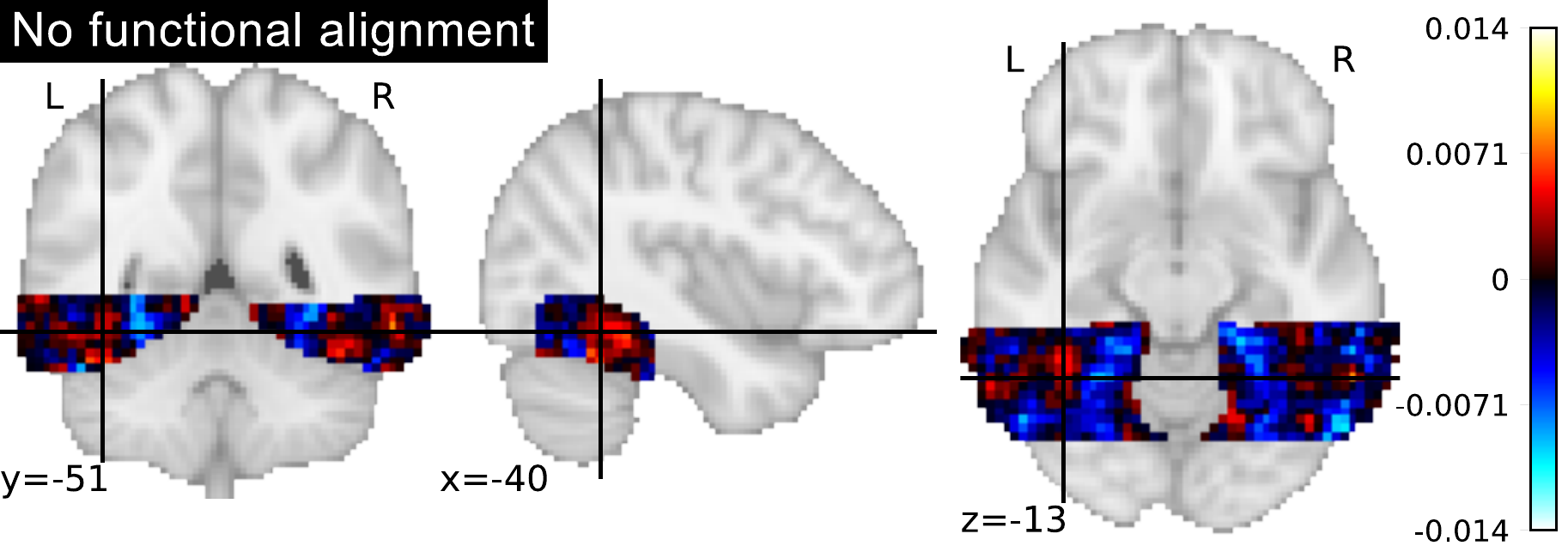}
	
	\endminipage\vfill
	\minipage{1\textwidth}
	\includegraphics[width=\linewidth]{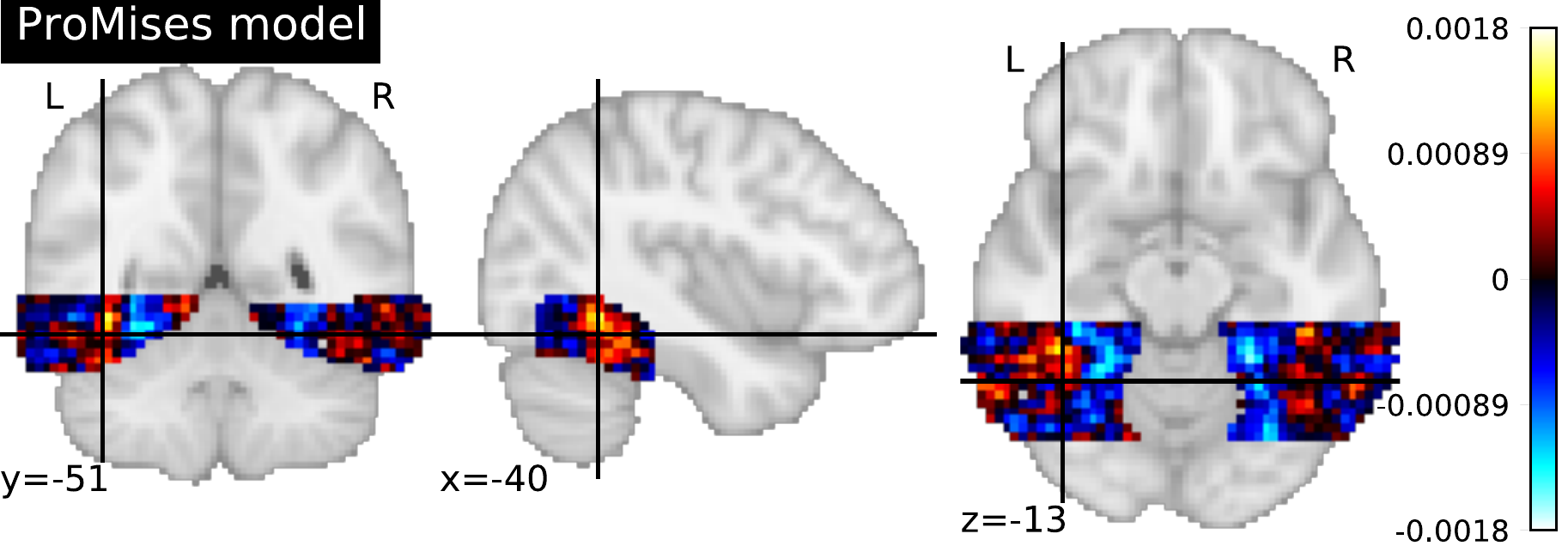}
	\endminipage
	\caption{Coefficients of the multi-class linear SVM considering the monkey face versus the male face classifier \angelaRev{(where hot colors correspond to predicting male face)} analyzing data aligned and not aligned via the ProMises model.}
	\label{fig:figsMale}
\end{figure}

We perform classification using the full set of voxels ($3509$), and plot the classifier coefficients in the brain space. With seven class categories and a one-versus-one strategy \citep{Lorena}, $21$ binary classifiers were fit. Figure \ref{fig:figsMale} represents the coefficients of the monkey face versus the male face classifier. The plots representing the coefficients of the classification of fine-grained distinctions in the object category and the coarse-grained distinctions between categories are shown in \ref{SVM}. 
In Figure \ref{fig:figsMale} we see that the coefficients of the classifier fit using anatomical alignment only is more diffuse than the equivalent values obtained using the ProMises model which appears to better capture the VT cortex's spatial anatomical geometry, as well as improve the ability to distinguish between categories.  

\angelaRev{The computation time equals $57.109$ seconds when no functional alignment is applied, while it equals $1619.835$ when using the ProMises model.}


\subsection{Visual object recognition}

\begin{figure}[!ht]
	\minipage{1\textwidth}
	\includegraphics[width=\linewidth]{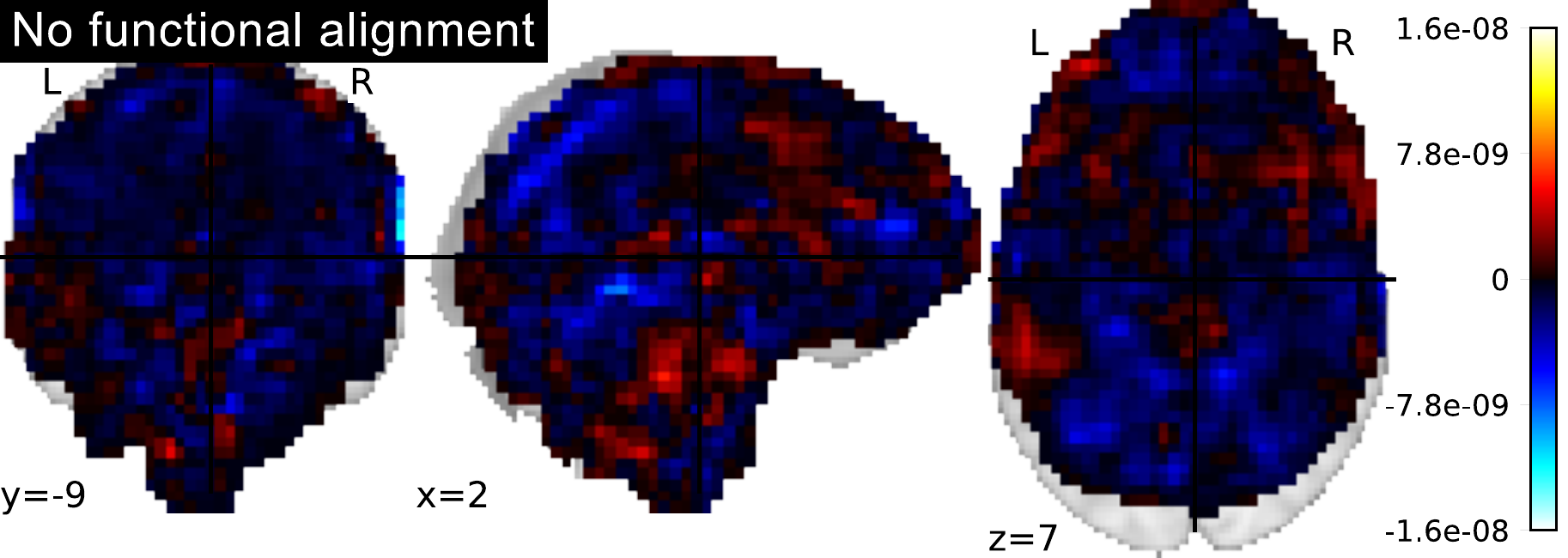}
	
	\endminipage\vfill
	\minipage{1\textwidth}
	\includegraphics[width=\linewidth]{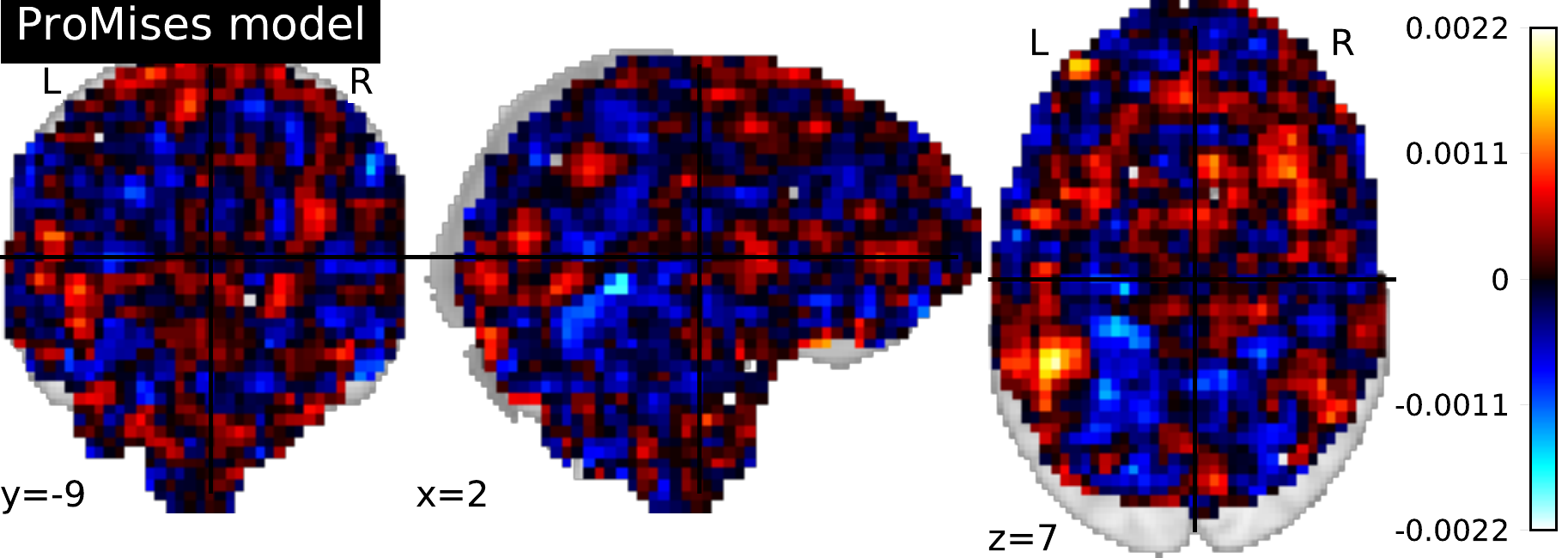}
	\endminipage
	\caption{Coefficients of the multi-class linear SVM considering the house versus face classifier \angelaRev{(where hot colors correspond to predicting human face)} analyzing data aligned  and not aligned via the ProMises model.}
	\label{fig:whole_brain}
\end{figure}

For this dataset the entire brain is functionally aligned using the Efficient ProMises model and classified via the SVM using the same process described for the \emph{faces and objects} dataset.    
Figure \ref{fig:whole_brain} represents the coefficients of the houses versus faces classifier using anatomically aligned only \citep{jenkinson2002improved} (top figure) and anatomically + functionally aligned (bottom figure) data. The Efficient ProMises model allows for the application of the classification to data from the entire brain, returning a between-subject accuracy equal to $0.6$, as well as a clear and interpretable brain map. In contrast, the anatomical alignment returns a more diffuse image with a between-subject accuracy equal to $0.4$.  

\begin{figure}[!ht]
	\includegraphics[width=\linewidth]{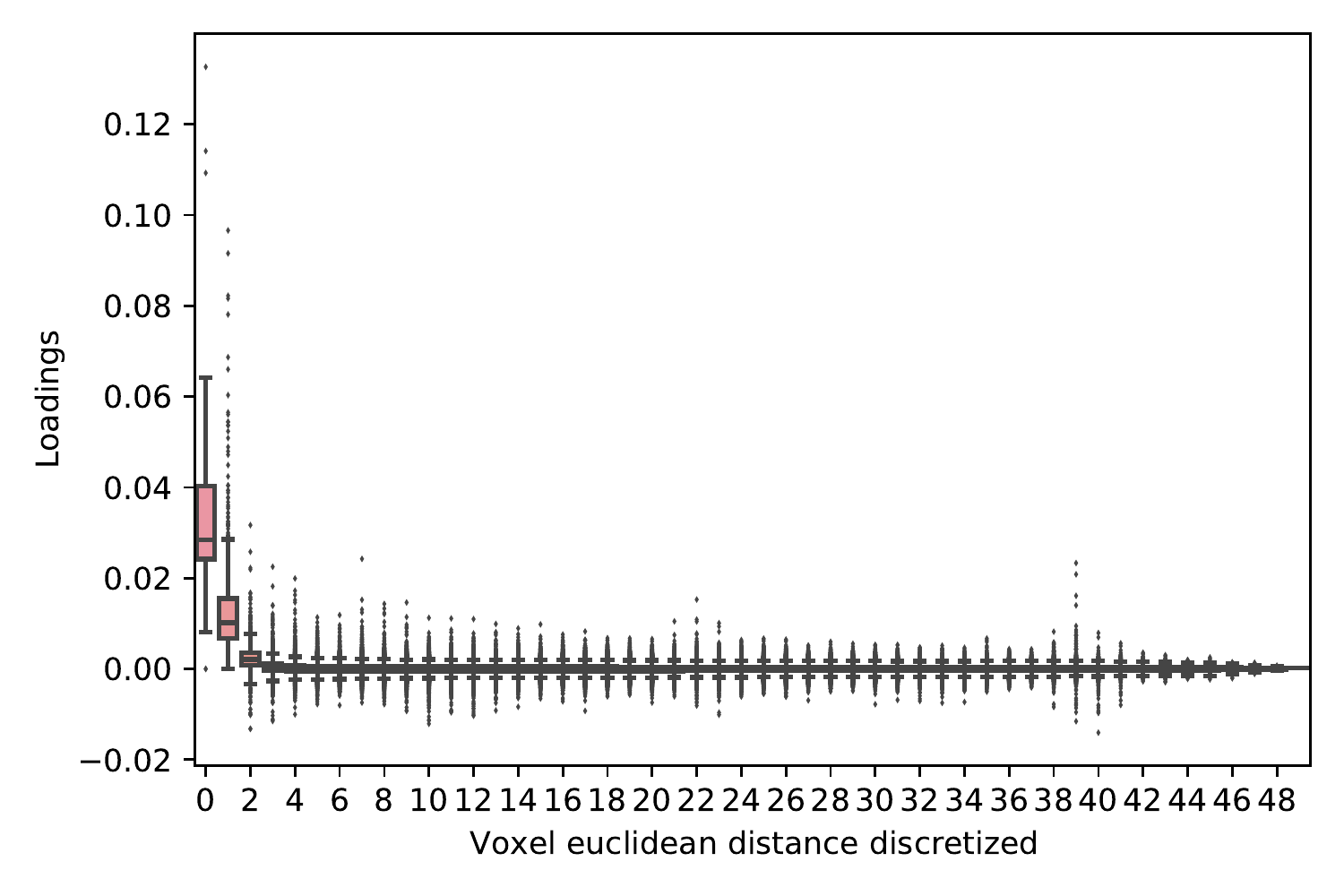}
	\caption{Boxplots of rotation loadings for each discretized value of the euclidean distance between $50$ voxels (randomly sampled) computed considering the $3D$ \angelaRev{voxel indices $ijk$} of the voxels. See the text for a detailed description of this figure.}
	\label{fig:loading_rot}
\end{figure}

One might think that whole-brain functional alignment is not recommended because idiosyncratic functional-anatomical correspondence generally occurs locally. The alignment must also avoid aligning different functional regions, such as the ventral temporal cortex of one subject with the prefrontal cortex in another subject. However, the Efficient ProMises model returns rotation coefficients that take into account the spatial brain information thanks to the specification of the prior distribution for the orthogonal parameters. These coefficients have high values for neighboring voxels and low values for distant voxels. For the \emph{visual object recognition} dataset, this result is shown in Figure \ref{fig:loading_rot} where the distribution of the loadings (i.e., contribution of the given voxel in the construction of the new, aligned, voxel) is shown as a function of the Euclidean distance ($3D$ \angelaRev{voxel indices $ijk$}) of the original voxels to the new voxel. For visualization purposes, the boxplots are grouped by the discretized value of the Euclidean distances. To clarify, let us consider as an example the first element $x_{11i}$ of the (non-aligned) matrix $\boldsymbol{X}_i$, and the first element $\hat{x}_{11i}$ of the aligned matrix $\hat{\boldsymbol{X}}_i$, where $\hat{x}_{11i} = x_{11i} r_{11i} +  x_{12i} r_{21i} + \dots + x_{1mi} r_{m1i}$ and $r_{kji} \in \boldsymbol{Q}_i\hat{\boldsymbol{R}}_i \boldsymbol{Q}_i^\top$. In Figure \ref{fig:loading_rot}, the voxel $x_{11i}$ will have a  distance equal to $0$ (in the abscissa), while the ordinate will be given by the value of the loading $r_{11i}$; other voxels $x_{12i}, \dots , x_{1mi}$ will have larger distances. Figure \ref{fig:loading_rot} shows that ProMises penalizes the combination of spatially distant voxels (i.e., loadings with small values) and prioritizes the combination of neighboring voxels (i.e., loadings with high values) in creating the new common abstract high-dimensional space.
\angelaRev{To further support this claim we report that the median cumulative proportion of squared loadings is 50\% at a distance of 19 voxels and is 90\% at a distance of 37.}
Thus, we claim that the Efficient ProMises model returns linear transformations for the whole brain that also act locally.
\angelaRev{The computation time equals $3554.375$ seconds if the ProMises model is used, while it equals $650.568$ seconds if no functional alignment is applied to the data.}

\subsection{Raiders}\label{raiders}

The voxel responses are from the VT, LO, and EV ROIs, which are essential brain regions for analyzing the subject's reaction to visual stimuli, such as watching a movie. The alignment and regularization parameter $k$  are computed using half of the movie and nested cross-validation, and the between-subject classification is performed on the remaining half to avoid circularity problems. The $1$-nearest neighbors algorithm is used to classify the correlation vector composed of six time points ($18$ second segment of the movie). The classification is correct when the correlation of the subject response vector with the group mean response vector (computed in the remaining subjects) is greater than the correlation between that vector response and the average response is to all other time segments. The classification is repeated for all $1$ hold-out subjects, and the average accuracy is computed as a performance metric.

The performance of the classification is tested using that was been anatomically alignned only, and data that is also functionally aligned using \emph{hyperalignment}, GPA, and the ProMises model. As we can see in Table \ref{tableAc1}, the ProMises model returns a higher mean accuracy than when only using anatomical alignment. The improvement in between-subject accuracy using the proposed method is consistent across the different ROIs and is roughly double of that obtained without any functional alignment.

\begin{figure}
    \minipage{0.32\textwidth}
    \includegraphics[width=\linewidth]{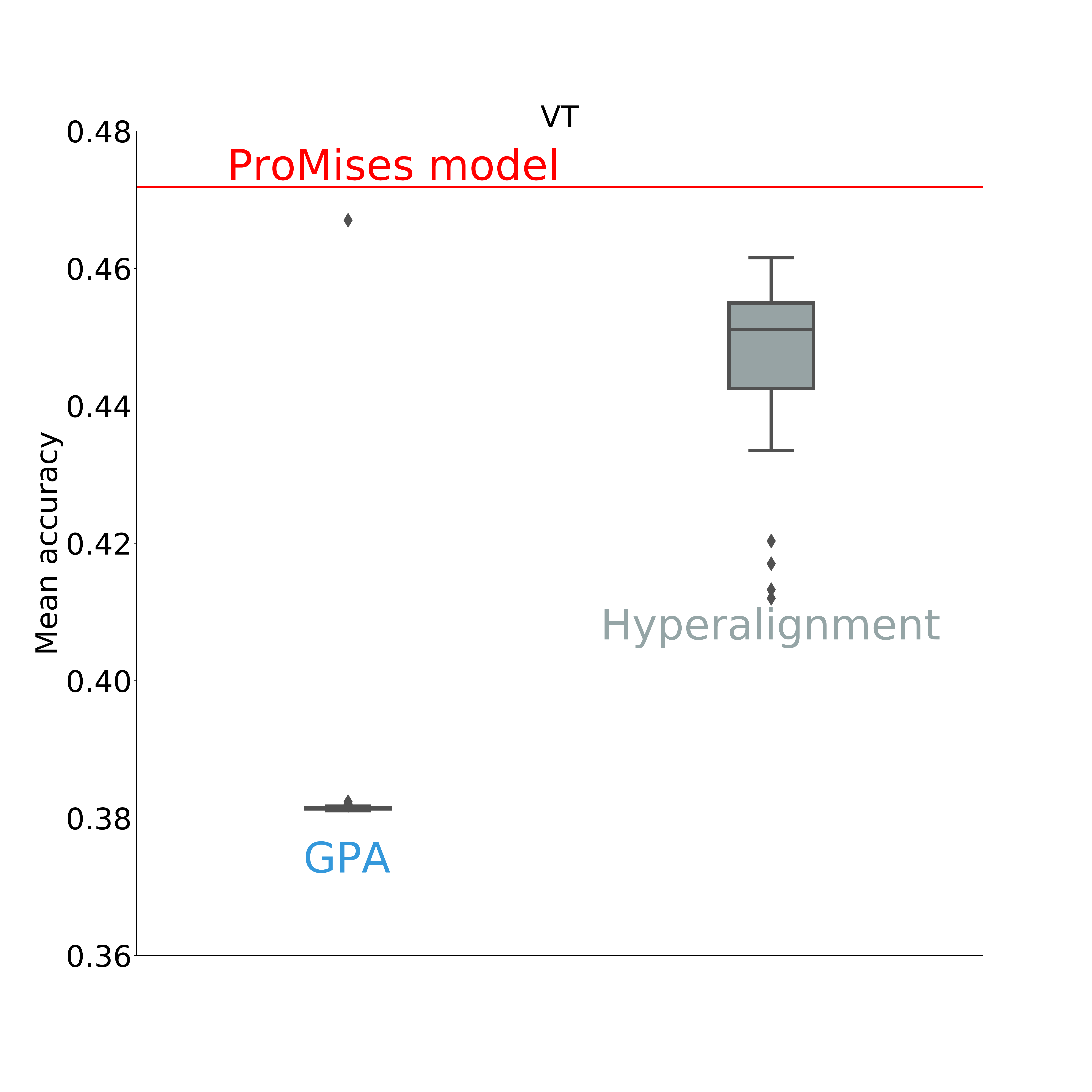}
    
    \endminipage\hfill
    \minipage{0.325\textwidth}
    \includegraphics[width=\linewidth]{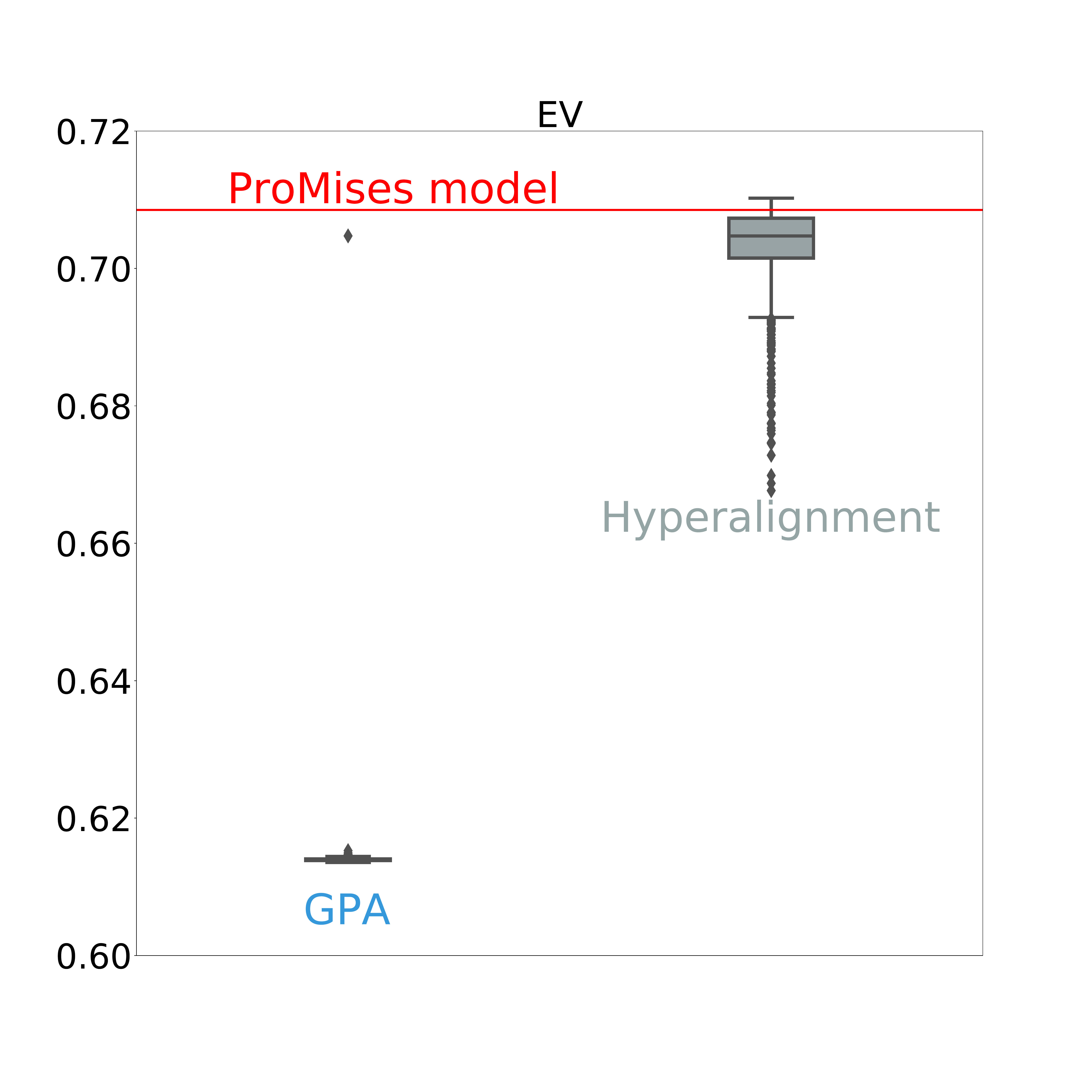}
    
    \endminipage\hfill
    \minipage{0.325\textwidth}%
    \includegraphics[width=\linewidth]{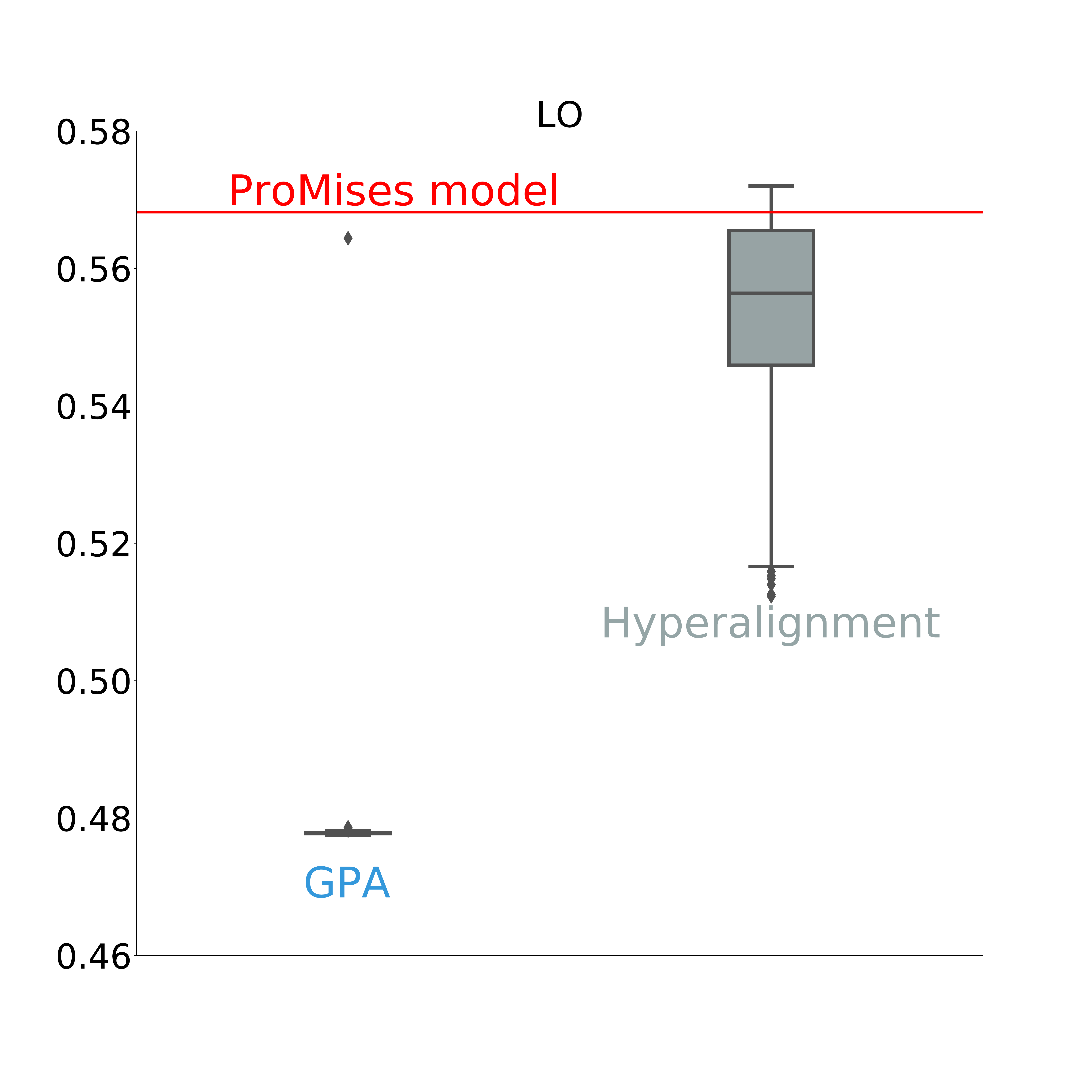}
    \endminipage
    \caption{Boxplots representing the mean classification accuracy for the \emph{raiders} data aligned using GPA and \emph{hyperalignment} for three different ROIs (VT, EV, and LO). The results obtained using the ProMises model are shown as a red line.
    }
    \label{fig:figsMovie}
\end{figure}

\angelaRev{It is important to note that various authors have already demonstrated this improvement in terms of classification accuracy using functional alignment as opposed to anatomical alignment \citep{Haxby}. However, here, we want to represent the variability of the between-subject accuracy if \emph{hyperalignment} or GPA are used instead of the ProMises model for functional alignment. In Figure \ref{fig:figsMovie} the gray boxplots represent the mean accuracy using \emph{hyperalignment}, having permuted the order of the subjects in the dataset $100$ times. In contrast, the blue boxplots show the mean accuracy of GPA using $100$ random rotations of the \emph{reference} matrix $\boldsymbol{M}$. Clearly, none of the two methods return a unique solution of $\boldsymbol{R}_i$, resulting in variability in the final classification results and complicating their interpretation. In contrast, the ProMises model provides a unique solution across all permutations, depicted \angelaRev{as} a red line in the figure.}

\angelaRev{Analyzing the \emph{hyperalignment} results, the variance of the between-subjects accuracy equals $0.0001$ for the VT analysis, $0.0002$ for the LO analysis, and $6.561\mathrm{e}{-5}$ for the EV analysis. In contrast, using GPA the variance equals  $0.0002$ for VT, $0.0001$ for LO, and $6.125\mathrm{e}{-5}$ for EV.} For all three ROIs, the accuracy obtained using the ProMises model is generally higher than the maximum values obtained using GPA. For \emph{hyperalignment}, the maximum value is higher than the results obtained using ProMises in EV and LO. Finally, similar to Section \ref{faces-and-objects}, the regularized \emph{hyperalignment} \citep{Xu} has the highest performance when defaulting to the standard \emph{hyperalignment} case for each ROI. 

\angelaRev{We also applied the \emph{regularized hyperalignment} approach proposed by \cite{Xu}; however, we found that the performance results are optimal with a regularized parameter equal to $1$ in all three frameworks (i.e., the \cite{Xu}'s method collapses to the standard \emph{hyperalignment} case).} 

\angelaRev{In the previous example, We empirically proved the non-uniqueness of \emph{hyperalignment} and GPA. For a formal proof, see \cite{Andreella}. This result means that we have a different representation of the aligned images and related results in the brain space for each set of transformations. However, using the ProMises model, this can be avoided. }

\begin{table}[]
\begin{tabular}{lllll}
&                                 & \multicolumn{3}{l}{\textbf{ROIs}}                                       \\ & \multicolumn{1}{l}{} & \multicolumn{1}{l}{VT}   & \multicolumn{1}{l}{EV}   & \multicolumn{1}{l}{LO}  \\ 
\cmidrule(l){2-5} 
\multicolumn{1}{l}{\multirow{2}{*}{\textbf{Alignment method}}} & \multicolumn{1}{l}{No functional alignment} & \multicolumn{1}{l}{$0.289$} & \multicolumn{1}{l}{$0.534$} & \multicolumn{1}{l}{$0.238$} \\ \cmidrule(l){2-5} 
\multicolumn{1}{l}{} & ProMises model & $0.472$ & $0.709$ & $0.568$ \\ \cmidrule(l){2-5} 
\end{tabular}
\caption{ Classification accuracy for the \emph{raiders} dataset using the anatomical alignment, as well as ProMises alignment for three different ROIs (VT, EV, and LO).}\label{tableAc1}
\end{table}

\angelaRev{Computation times are reported in Table \ref{time_raiders} for each analysis.}

\begin{table}[]
\begin{adjustbox}{width=\columnwidth,center}
    \centering
\begin{tabular}{llll}                  \\ \multicolumn{1}{l}{} & \multicolumn{1}{l}{VT}   & \multicolumn{1}{l}{EV}   & \multicolumn{1}{l}{LO}   \\ 
                              \toprule
                                      \rowcolor{Gray}

\textbf{No functional alignment} & $234.32$ & $204.798$  & $188.003$ \\ 
\textbf{ProMises model} & $7581.615$ & $7003.431$ & $7423.76$ \\
        \rowcolor{Gray}

\textbf{Hyperalignment} & $2728.47$ & $2102.53$ & $2481.74$ \\
\textbf{GPA} & $48847.36$ & $41748.33$ & $47890.32$ \\
\end{tabular}
\end{adjustbox}
\angelaRev{\caption{Computation time (in seconds) for each analysis performed on the \emph{raiders} dataset.}\label{time_raiders}}
\end{table}

\subsection{Words and objects}\label{words_objects}
\begin{figure}[!ht]
    
    \minipage{1\textwidth}
    \includegraphics[width=\linewidth]{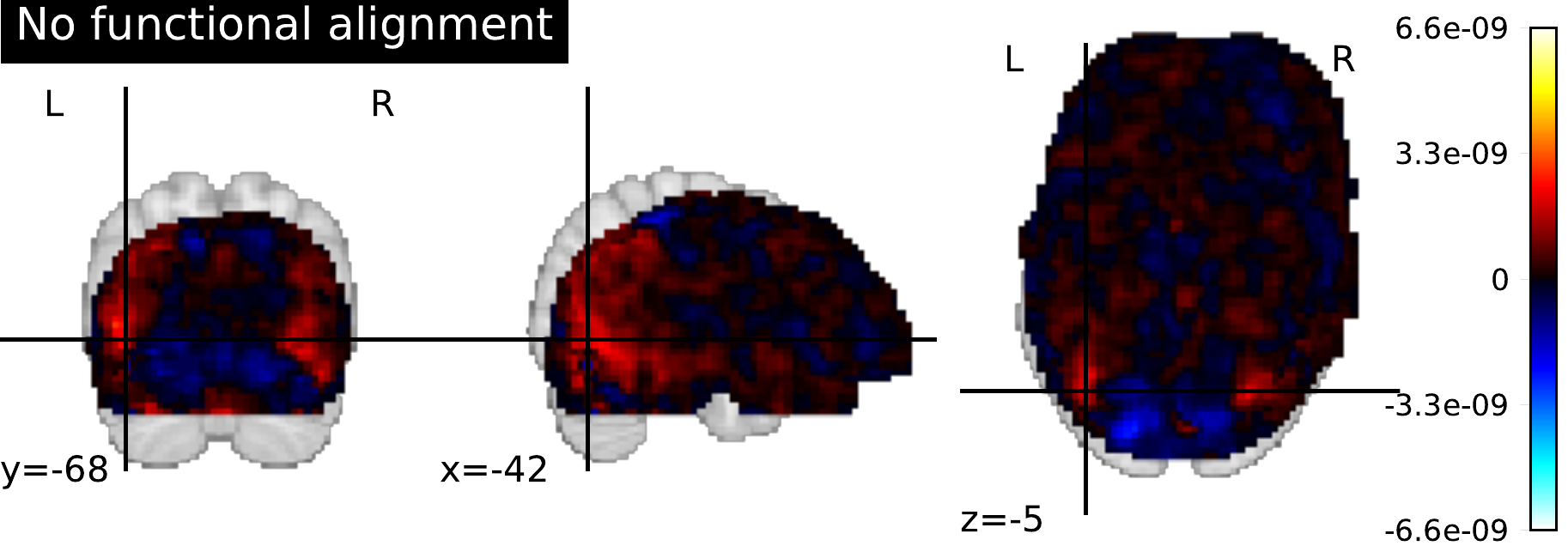}
    
    \endminipage\vfill
    \minipage{1\textwidth}%
    \includegraphics[width=\linewidth]{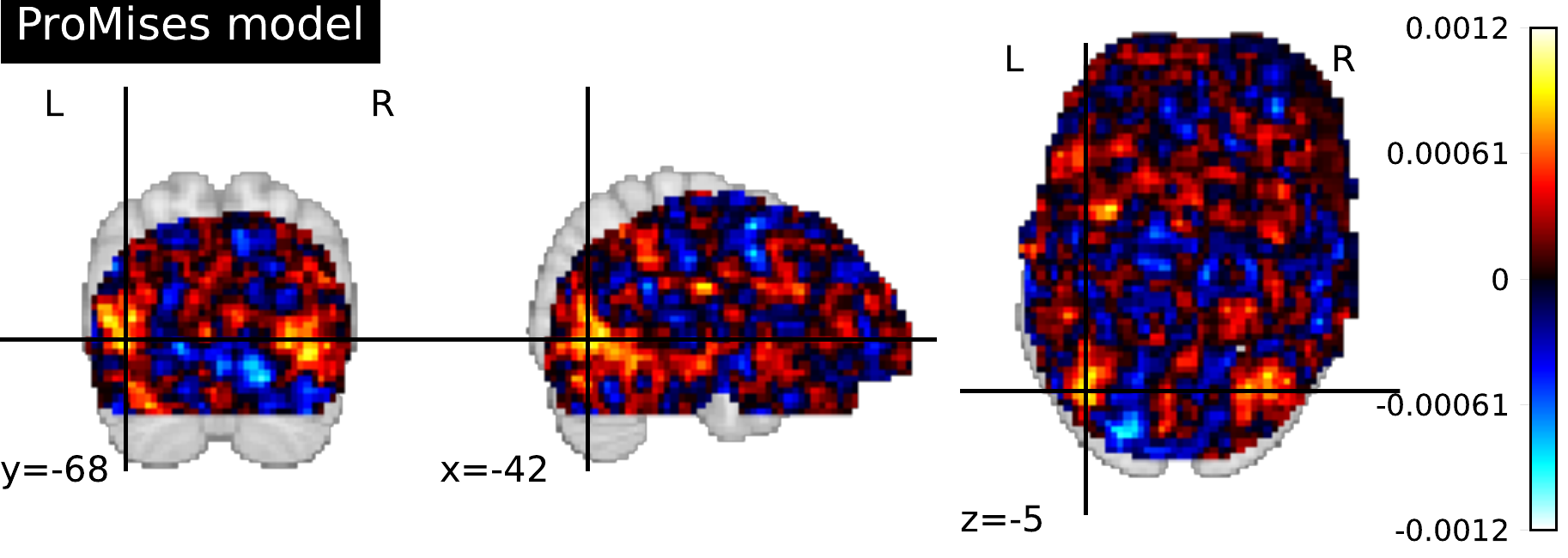}
    \endminipage
    
    \caption{\angelaRev{Coefficients of the multi-class linear SVM considering the consonant string versus scrambled objects classifier (where hot colors correspond to predicting consonant string) analyzing data aligned via anatomical alignment and the ProMises model.}}
    \label{fig:words}
\end{figure}

\angelaRev{In this analysis, the entire brain is functionally aligned using the Efficient ProMises model in the same manner as described in the \emph{visual objects recognition} data analysis. The brain images are then classified via SVM following the same procedure used for the \emph{faces and objects} and \emph{visual objects recognition} datasets. Figure \ref{fig:words} shows the coefficients of the consonant string versus scrambled objects classifier considering functionally aligned (bottom figure) and not functionally aligned (top figure) data. The between-subject accuracy equals $0.2$ if the data is not functionally aligned, while it is $0.33$ if the data is functionally aligned. It is interesting to note the two yellow blobs which correspond to Brodmann area $19$, which is known to be a visual processing area \citep{wright2008selective, duncan2009consistency}, where heightened activation corresponds to predicting the consonant string.  The analysis takes $6557.867$ seconds if the Efficient ProMises model is used, in comparison it takes $1204.945$ seconds if it is not used.}

\section{Discussion}

Functional alignment is a preprocessing step that improves the functional coherence of fMRI data, hence improving the accuracy of subsequent analysis \citep{haxby2020hyperalignment, Haxby, Guntupalli}. 

This paper presents a functional alignment method that solves most of the shortcomings present in previously used methods, particularly GPA \citep{GowerGPA} and \emph{hyperalignment} \citep{Haxby}.
The ProMises model returns aligned images that are interpretable, fully reproducible, and provide enhanced detection power.
Below we summarize several of the key findings of this paper.

\subsection{Enhanced detection power}

We applied the proposed method to \angelaRev{four} different data sets, allowing us to evaluate its performance under a number of different settings. This included differences in sample size, length of time series, whether data was extracted from an ROI or the whole-brain, and whether the data resides in volumetric space or on the surface.

We further contrasted the approach with three other approaches. The first was simply performing no functional alignment (i.e., anatomical alignment only). The second, was the standard \emph{hyperalignment} approach. The third, was the classic GPA approach. For the two latter approaches and ProMises anatomical alignment was performed prior to functional alignment.

For all \angelaRev{four} datasets the ProMises model greatly outperformed using no functional alignment\angelaRev{, e.g., see Table } \ref{tableAc1}. Consistently for all settings the classification accuracy was roughly doubled when using the ProMises model in addition to standard anatomical alignment. Further, as can be seen in Figure \ref{fig:figsMovie}, in most cases, the ProMises model outperformed the other functional alignment techniques in {\em all} permutations of these approaches. On occasion certain permutations outperformed ProMises, though this was rare. 

\subsection{Reproducibility and interpretability of the results}

The solution (i.e., the final image) produced by GPA is not unique, as it depends on the starting point of the iterative algorithm. A similar issue arises for the \emph{hyperalignment} method where the solution depends on the order in which subjects are entered into the algorithm. As a consequence, results may vary widely depending on arbitrary choices made by the experimenter (or the software used). The severity of this problem is visible in all analyses performed in this paper. As an example see Figure \ref{fig:figsMovie}, where the accuracy is not given by a single value, but rather represented using a box-and-whisker plot. To the best of our knowledge, the
ProMises model is the only \angelaRev{Procrustes-based} functional alignment technique that resolves the problem of non-uniqueness of the solutions, thereby enhancing the reproducibility of the results.

The non-uniqueness of the solution further leads to difficulties with interpretability, since equivalent solutions in mathematical terms (i.e. where the same maximum is obtained)
may provide different -- sometime very different -- final images. This reduces the previous alignment methods to black box solutions that do not directly improve our understanding of the underlying cognitive activities.

The ProMises model offers a way to address these two issues thanks to the inclusion of prior (and anatomical) information into the analysis. This makes the solution unique (i.e. reproducible), and the resulting images interpretable.
The proposed solution borrows information from the whole brain, but is driven to act locally, therefore making the results anatomically meaningful. This can clearly be seen in Figure \ref{fig:loading_rot}, which reports the contribution of a given voxel in the construction of the new, aligned, voxel as a function of the Euclidean distance. It is evident that the highest contribution comes from the voxels that are closest in proximity.

As confirmation of the interpretative quality of the method, we can study the map of classifier coefficients presented in Figure \ref{fig:whole_brain}. The image is clear and interpretable if the functionally aligned fMRI data are used. For example, we can see a yellow blob of activation in Figure \ref{fig:whole_brain} in the functionally aligned data. The blob corresponds to the superior temporal gyrus, a region known to be involved in the perception of emotions in reaction to facial stimuli \citep{haxby2001distributed, ishai2000representation, ishai1999distributed}. While, we can comfortably interpret these maps when using the ProMises model, it is more ambiguous for other methods. In fact, the other methods do not return a single aligned image; the representation of the results on the anatomical template is possible but without guarantee of validity from a mathematical point of view. \angelaRev{We stress here that the classifier weight coefficients must be transformed to proper activation patterns \citep{haufe2014interpretation} if inferential conclusions are desired.}

\subsection{Computationally efficiency}

While the proposed method is iterative, it is usually less computationally intensive than GPA (i.e. the non regularized counterpart). The reason is that it typically reaches the convergence criteria in only a few iterations thanks to the regularization term defined by the prior parameters (i.e.  $k$ and $\mathbf{F}$).
As an example, consider the first analysis in Subsection \angelaRev{\ref{raiders}} (i.e., using \angelaRev{the VT mask from the \emph{raiders}} dataset). Here, the ProMises model takes \angelaRev{$7581.615$} seconds to perform the analysis, whereas GPA takes \angelaRev{$48847.36$} seconds using a $3000+$ core Linux cluster with $20$ GB of random access memory and parallel computation for the subjects \angelaRev{(i.e., the analysis are parallalized across a number of cores equal to the number of subjects included in each analysis)}.  \emph{Hyperalignment} only takes \angelaRev{$2728.47$} seconds, but it is does not reach any optimality criterion, as seen in Section \ref{Procrustes}.         \angelaRev{Finally, the computation time could be improved by using different approaches than cross-validation (e.g., generalized cross validation \citep{golub1997generalized} or bandwidth selection techniques \citep{heidenreich2013bandwidth}).}

\subsection{Whole brain applicability}

More relevantly, the efficient extension of the  ProMises model overcomes computational difficulties related to performing whole-brain analysis that plague both \emph{hyperalignment} and GPA. This extension of the model works on a reduced space of the data, thereby gaining in efficiency. In practice, the dimensions are reduced from the number of voxels to the number of scans, which for typical fMRI data implies a significant dimension reduction.
A competing model in this context is \emph{searchlight hyperalignment} \citep{Guntupalli}, where overlapping transformations are calculated for overlapping searchlights in each subject and then aggregated into a single whole-brain transformation. While this allows for an anatomical interpretation of the final map, the final transformation is not an orthogonal matrix, and therefore will not preserve the content of the original data. While \emph{searchlight hyperalignment} uses local radial constraints, the ProMises model incorporates them directly into the Procrustes estimation process through the prior, thus providing increased flexibility. \angelaRev{Another approach is \emph{piecewise} functional alignment \citep{bazeille2021empirical}, where non-overlapping regions (coming from a priori functional atlas or parcellation methods) are aligned and then  aggregated. \cite{bazeille2021empirical} found substantial improvement compared to using \emph{searchlight} approaches in whole brain analysis. However, it suffers from possible staircase effects along the boundaries of the non-overlapping regions.}

\subsection{Extensions}
Because the proposed approach is a statistical model, various extensions can be considered to include more flexibility (e.g., examining sub-populations using different reference or location matrices). The specification of location matrix $F$ as a similarity matrix also permits exploring various types of distances (e.g., considering the gyrus instead of the voxels as units). To conclude, the definition of $F$ opens up a universe of different possibilities to express anatomical and functional constraints existing between voxels/regions in the brain. \angelaRev{It is plausible that other functional alignment methods proposed in the literature (e.g., SRM proposed by  \cite{Xu}) can be incorporated into the ProMises model, which can be explored in future work.}

\section{Conclusion}

Together, these findings lead us to believe that the ProMises algorithm provides a promising approach towards performing functional alignment on fMRI data that improves classification accuracy across a number of different settings. We therefore believe it is an attractive option for performing functional alignment on the fMRI data prior to fitting predictive models.



\section*{Code availability}
The custom code used for the analysis can be downloaded from \url{https://github.com/angeella/ProMisesModel}.

\section*{Data availability}
\angelaRev{The faces and objects dataset can be downloaded from \url{https://github.com/angeella/ProMisesModel/tree/master/Data/Faces_Objects}, while the visual object recognition dataset from \url{https://www.openfmri.org/dataset/ds000105/}. The raiders dataset is currently not available online, but data can be made available upon request. Finally, the \emph{words and objects} dataset can be downloaded from \url{https://www.openfmri.org/dataset/ds000107/}.}

\section*{CRediT authorship contribution statement}
\textbf{Angela Andreella}: Conceptualization, methodology, formal analysis, and writing of the original draft. \textbf{Livio Finos}: Conceptualization, methodology, writing the original draft, and supervision.
\textbf{Martin A Lindquist}: Conceptualization,  writing the original draft, and supervision.

\section*{Declaration of Competing Interest}
The authors declare no competing interests.

\section*{Acknowledgements}
We would like to thank Prof. James Van Loan Haxby, Prof. Yaroslav Halchenko, and Dr. Ma Feilong for sharing the fMRI data used in this manuscript, for interesting insights and comments, and for hosting Angela Andreella at the Center for Cognitive Neuroscience at Dartmouth College. Some of the computational analyses done in this manuscript were carried out using Discovery Cluster at Dartmouth College (\url{https://rc.dartmouth.edu/index.php/discovery-overview/}). Angela Andreella gratefully acknowledges funding from the grant BIRD2020/SCAR\_ASEGNIBIRD2020\_01 of the University of Padova, Italy, and PON 2014-2020/DM 1062 of the Ca’ Foscari University of Venice, Italy. Martin A Lindquist was supported in part by NIH grants R01 EB016061 and R01 EB026549 from the National Institute of Biomedical Imaging and Bioengineering.

\clearpage
\bibliographystyle{apalike}
\bibliography{bibliography}
\clearpage

\appendix

\section{Algorithms}\label{pseudocode}

Algorithm \ref{alg:algo} provides the pseudocode for the estimation process of the ProMises model, while Algorithm \ref{alg:algo1} shows the adaptation to perform its Efficient version.

\begin{algorithm}
	\begin{algorithmic}[1]
		
	\Require $\boldsymbol{X}_i$, $\mathtt{T}$ , $\mathtt{maxIt}$, $\forall i=1,\dots,m$
	\Ensure $\hat{\boldsymbol{X}}_i$ $\forall i=1,\dots,m$
\State	$\boldsymbol{\hat{M}} = \sum_{i = 1}^{m} \boldsymbol{X}_i /m$
\State	$\mathtt{count} = 0$, $\mathtt{dist} = \mathtt{Inf}$
	\While{$\mathtt{dist}$ $>$ $\mathtt{T}$  OR  $\mathtt{count}$ $<$ $\mathtt{maxIt}$}
		\For{$i=1$ to $m$}
		\State	$\boldsymbol{U}_i \boldsymbol{D}_i \boldsymbol{V}_i^\top = \texttt{SVD}(\boldsymbol{X_i}^\top  \boldsymbol{\hat{M}}  + k \boldsymbol{F})$ \Comment{\small Singular value decomposition } \label{alg:svd}
		\State	$\boldsymbol{\hat{R}}_i = \boldsymbol{U}_i \boldsymbol{V}_i^\top$
		\State	$\boldsymbol{\hat{X}}_i = \boldsymbol{X}_i \boldsymbol{\hat{R}}_i$ \Comment{\small Update $\boldsymbol{X}_i$}
		
	\EndFor
	\State	$\boldsymbol{\hat{M}}_{\text{old}} = \boldsymbol{\hat{M}}$\Comment{\small Save $\boldsymbol{\hat{M}}$}
	\State	$\boldsymbol{\hat{M}}=\sum_{i = 1}^{m}\boldsymbol{\hat{X}}_i / m$\Comment{\small Update $\boldsymbol{\hat{M}}$}
	\State	$\mathtt{dist} = ||\boldsymbol{\hat{M}} - \boldsymbol{\hat{M}}_{\text{old}}||^2$, $\mathtt{count} = \mathtt{count} + 1$.
	\EndWhile
	
	\caption{ProMises model algorithm, where $\mathtt{T}$ is the threshold for $||\hat{\boldsymbol{M}} - \hat{\boldsymbol{M}}_{old}||^2$, $\mathtt{maxIt}$ is the maximum number of iterations allowed, and $\epsilon_1$ and $\epsilon_2$ denote two infinitesimal positive quantities.}\label{alg:algo}
\end{algorithmic}
\end{algorithm}

\begin{algorithm}[H]

	\caption{Efficient ProMises model algorithm.
	}\label{alg:algo1}
	
	Use Algorithm \ref{alg:algo}, only add these three lines at the beggining:
	\begin{algorithmic}[1]
		\For{$i=1$ to $m$}
		\State $\boldsymbol{L}_i \boldsymbol{S}_i \boldsymbol{Q}_i^\top = \texttt{SVD}(\boldsymbol{X_i})$ \Comment{\small Thin singular value decomposition}	
		\State $\boldsymbol{X}_i  = \boldsymbol{X_i} \boldsymbol{Q}_i$
		\EndFor
\end{algorithmic}
\end{algorithm}

\section{Faces and objects results}\label{SVM}
\begin{figure}[!ht]
    \minipage{1\textwidth}
    \includegraphics[width=\linewidth]{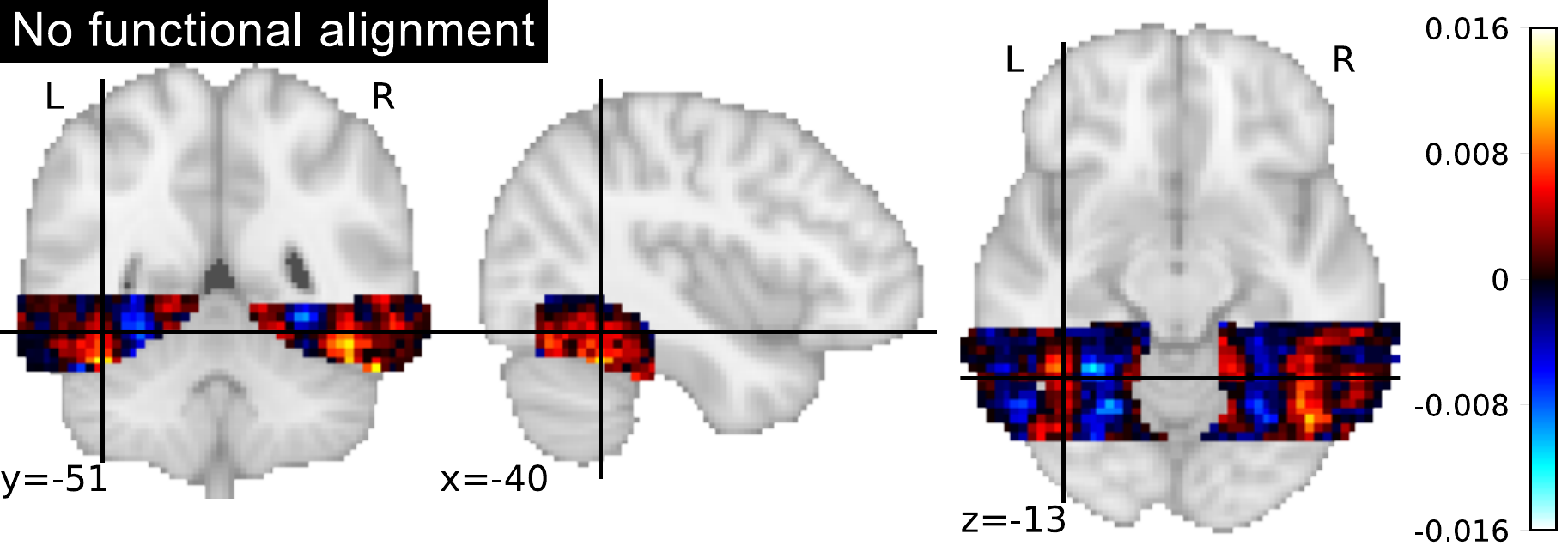}
    
    \endminipage\vfill
   
    \minipage{1\textwidth}%
    \includegraphics[width=\linewidth]{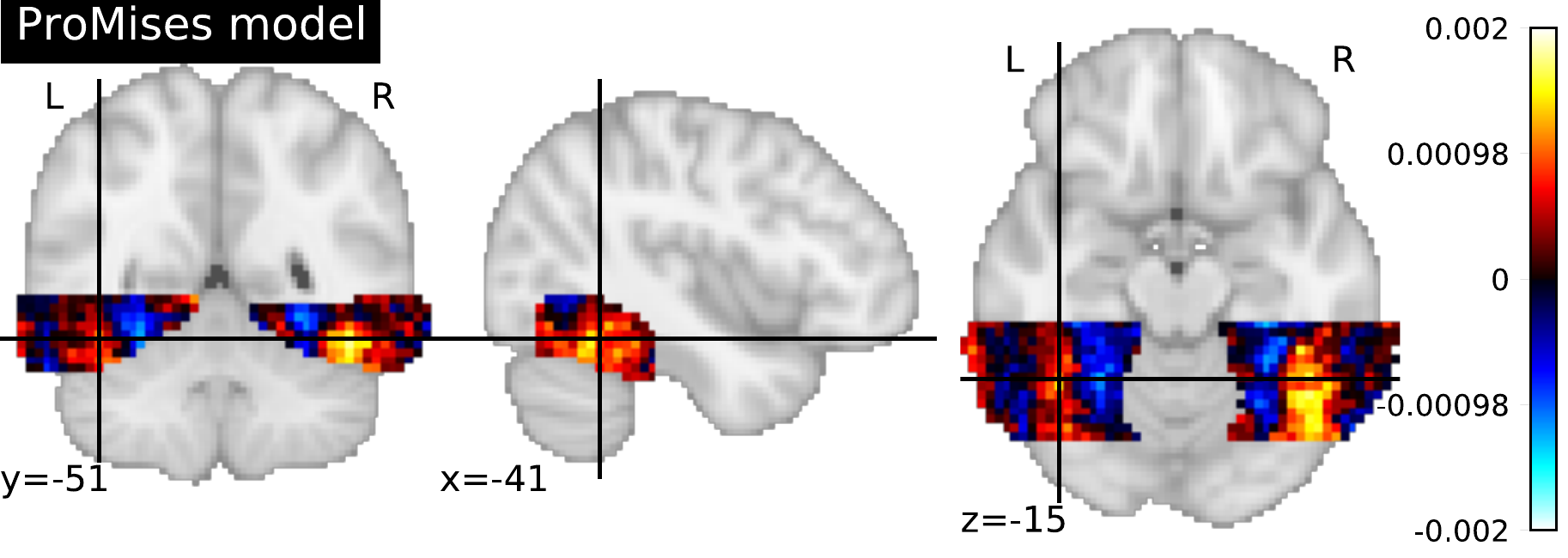}
    \endminipage\vfill
    
    \caption{Coefficients of the multi-class linear SVM considering the dog face versus shoe classifier \angelaRev{(where hot colors correspond to predicting dog face)} analyzing data aligned via anatomical alignment and the ProMises model.}
    \label{fig:figObj}
\end{figure}

\begin{figure}[!ht]
    
    \minipage{1\textwidth}
    \includegraphics[width=\linewidth]{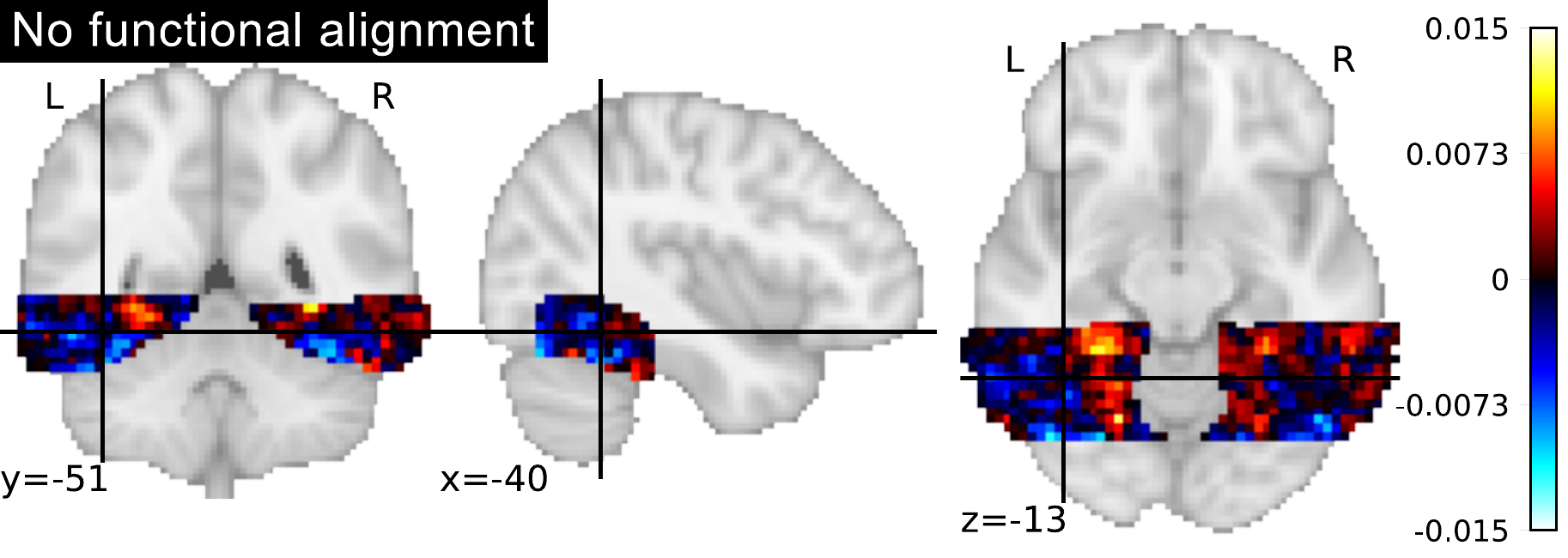}
    
    \endminipage\vfill
    \minipage{1\textwidth}%
    \includegraphics[width=\linewidth]{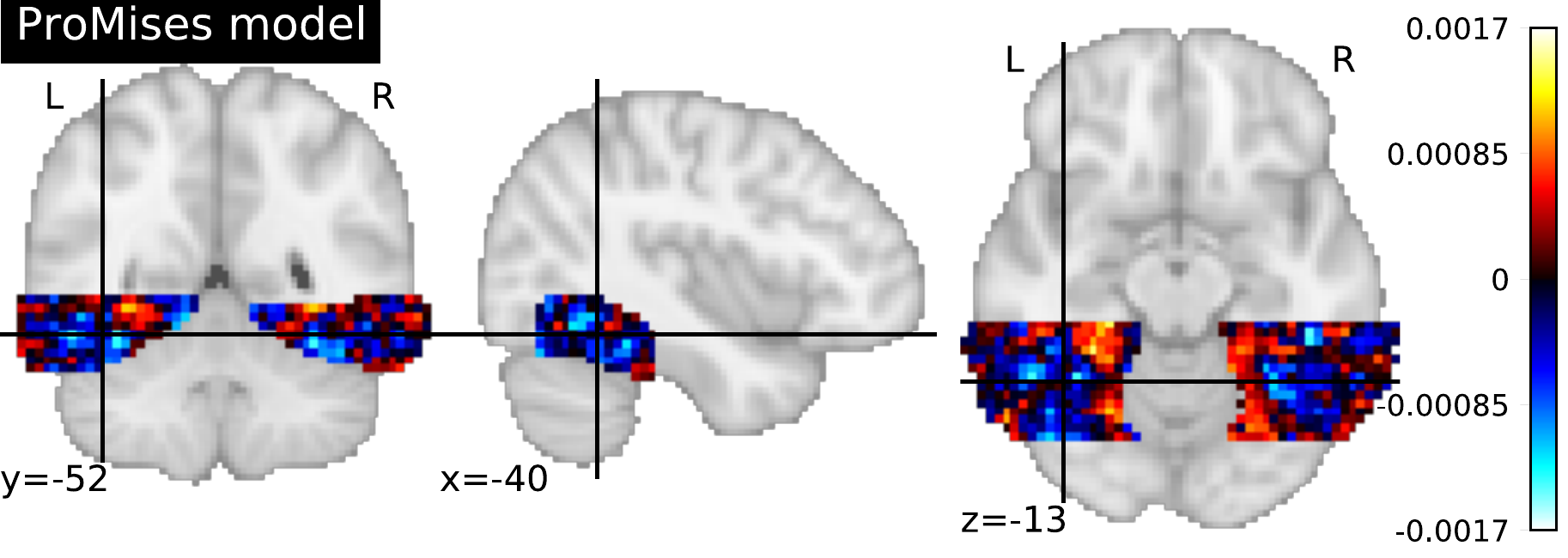}
    \endminipage
    
    \caption{Coefficients of the multi-class linear SVM considering the chair versus shoe classifier \angelaRev{(where hot colors correspond to predicting chair)} analyzing data aligned via anatomical alignment and the ProMises model.}
    \label{fig:figCoarse}
\end{figure}

Figure \ref{fig:figObj} represents the coefficients considering a classification of fine-grained distinction among object categories (i.e., shoes versus a chair). In the same way, Figure \ref{fig:figCoarse} represents the case of coarse-grained distinctions (i.e., dog face versus shoe). Thanks to the model formulation of the proposed method, the coefficients of the classifiers can be represented in brain space, returning maps with spatial boundaries between different categories of stimuli.

\clearpage





\end{document}